# {110} Plane Orientation Driven Superior Li-ion Battery Performance and Room Temperature Ferromagnetism in $Co_3O_4$ Nanostructures

R. Zeng[†,‡], J.Q. Wang[†,∥], G.D. Du[†,‡], W.X. Li[†], Z.X. Chen[†], S. Li[§], Z.P. Guo[†], S.X. Dou[†]

[†]Institute for Superconducting and Electronic Materials, School of Mechanical, Materials & Mechatronics Engineering, University of Wollongong, NSW 2522, Australia.

[‡]Solar Energy Technologies, School of Computing, Engineering and Mathematics, University of Western Sydney, Penrith Sout, Sydney, NSW 2751, Australia.

[∥]School of Materials Science and Engineering, University of Jinan, Jinan 250022, P. R. China.

[§]School of Materials Science and Engineering, University of New South Wales, Sydney NSW 2502, Australia.

Address for Correspondence:

R. Zeng

Solar Energy Technologies
School of Computing, Engineering and Mathematics
University of Western Sydney
Penrith Sout, Sydney, NSW 2751, Australia
Electronic mail: R.Zeng@uws.edu.au

# ABSTRACT

This paper presents synthesis methods and multifunctional properties of porous $Co_3O_4$ nanoplatelets and nanowires. Surprisingly, significantly superior Li-ion battery performance compared to other cathode materials is exhibited in the same samples as room temperature ferromagnetism (RTFM). Microstructure observations, and properties measurements and analysis indicate that the as-prepared $Co_3O_4$ nanostructures exhibit a spontaneous transformation of their morphology, showing a predominantly {110} plane orientation and preferentially presenting the $Co^{3+}$ species at the surface at appreciate preparation condition. More $Co^{3+}$ in an octahedral (***O***) position exposed on the surface is associated with more $Co^{3+}$ being $Co^{3+}|\bullet$ ($Co^{3+}$ with trapped charge carriers) or reduced to $Co^{2+}$ in the ***O*** position due to surface defects or oxygen vacancies, which is possibly the mechanism of excellent electric transport properties then high Li-battery performance; at the same time, reduced to $Co^{2+}$ or free charge carriers trapped by $Co^{3+}|\bullet$ in the ***O*** position are the role of charge carries-mediate ferromagnetism, may be the origin of RTFM as well. These results further show the importance of modifying the nanostructure of surfaces in the preparation of transition metal oxides as multidisciplinary functional materials.

## 1. Introduction

Transition metal oxides play an important role in a variety of technological applications and heterogeneous processes. Among the most extensively studied oxides, $Co_3O_4$, is arousing increasing interest in multidisciplinary fields, such as energy storage - Li-ion batteries [1-4], and environmental science - catalysts [5, 6]. Doped $Co_3O_4$ is also of interest as a multferroic material [7] and as a parent compound for other layer-structured cobalt oxides, e.g. $Na_xCoO_2$ in the thermopower field [8] and as a superconductor [9] Understanding of metal–oxide nanostructures in-depth is crucial to continued development of the technological performance and efficiency of $Co_3O_4$-based materials multifunctional applications.

The Li-ion battery is the main work horse for present and future energy storage devices Its storage of electrical energy at high charge and discharge rates is an important technology that enables hybrid and plug-in hybrid electric vehicles and provides backup for wind and solar energy [10]. It is believed that in the Li-ion battery system, very high power density and rates may be very hard to achieve concurrently, due to electron and charge transport limitations, which involve a trade-off of high power for low energy density, as they only store energy by surface adsorption reactions of charged species on an electrode material [11-14]. On the other hand, $Co_3O_4$ is an example of an important *p*-type semiconductor with direct optical band gaps at 1.48 and 2.19 eV [15]. In addition to its magnetic behaviour, it is also a very important magnetic semiconductor [16, 17] and multiferroic material [7] for multifunctional electronic devices and information storage applications. Most recently, the electronic states and magnetic structure at $Co_3O_4$ (110) surface have been studied by a first principles [18], it is indicated that, the $Co^{3+}$ ions do not have a magnetic moment in the bulk, but become magnetic at the surface, which leads to surface magnetic orderings different from the one in the bulk. Surface electronic states are present in the lower half of the bulk band gap and cause partial metallization of (110) surface terminations. This imply the complicated of conductivity and magnetism of (110) surface state. The better conductivity enhancement is very benefit to the Li-ion battery performance – improve the electronic transport properties; at the same time, the magnetic properties are very sensitive to the electron state, band-structures, and defects in the materials, the magnetic properties of bulk materials are normally well-defined, and most possible changes is the surface states and properties since the large surface area / volume rate in nanostructures. So the variation of magnetic properties in nanomaterial does most likely directly reflect to / related to the changes of the surface electron, defects or disorder states, and interaction between surface layers with bulk core. It is an effective means for employing the magnetic measurements to evaluate the electron states (free / unpaired / localised), band energy states / structure in different nanostructures. For example, from the temperature dependence curve of susceptibility, we can distinguish the free electron and localised electron spin contributions [19], which direct associate with conductivity behaviours of nanostructures. Moreover, the Van Vleck susceptibility is inversely proportional to band gap $E_g$ [20, 21] from the Van Vleck susceptibility results, it can be evaluated that the band gap changes with temperature, particle size and surface area / volume rate of different nanostructures. Since it is very difficult to evaluate the electron / charge transport properties of nanostructures used for electrochemical or photoelectrochemical cells, the magnetic measurement is very efficiency method. Currently, researchers measure the conductivity of single nanowires by DC I-V method, and STEM method. The problems with these special measurements are that (1) these special tests need unique facility and proper device setup, and are very difficult to conduct, (2) the results by these method only reflects very limited features on few nanowires or small areas. We noticed that there are large number of reports deal with the links between magnetic properties with Li-ion battery materials [22] They reviewed the fundamental magnetic studies on well structurally characterized standard samples for understanding of Li-ion battery materials with a notion to estimate the non-stoichiometry, to monitor sample purity. When the structural defects appearing due to disorder, nano-sizing etc, the magnetic studies should be corroborated to other technique. We have developed a magnetic susceptibility measurements method to evaluate the electron / charge transport characteristics in different nanostructures. Since the magnetic susceptibility reflect free electron and unpaired electron features, and it is the average signal from few ten grams powder, the magnetic susceptibility measurement can be used to characterize the electron/charge transport in the large scale nanomaterials. The magnetic susceptibility measurement have been successfully used in investigating the charge transfer behaviour in lithium ion batteries previously [23, 25]. Here will continue to rich this technique to analysis the charge transport properties combining the electrochemical measurements in these $Co_3O_4$ unique nanostructures.

Here, we report the preparation, microstructure and electron structure, electrochemistry / Li-ion battery and magnetic properties of $Co_3O_4$ mesoporous nanoplatelets and nanowires. Systematic comparison and analysis of the relationships among the grain size, microstructure, orientation (texture), and surface defects with both electrochemistry / Li-ion battery and magnetic properties are presented in this work.What is more surprising is the significantly superior Li-ion battery performance (where the batteries concurrently present high power density and long cycle life at high energy density rates) compared to other cathode materials for the same samples where room temperature ferromagnetism occurs, which further shows the importance of modifying nanostructure surfaces in the preparation of transition metal oxides as multidisciplinary functional materials.

## 2. Experimental

**Materials preparation and characterization**

All the reagents used in the experiment were analytically pure. First, cobalt hydroxide nanoplatelets were prepared by a precipitation and microwave hydrothermal process. The experimental details are as follows: 10 mmol $Co(NO_3)_2 \cdot 6H_2O$ was dissolved in 50 ml distilled water under stirring. Then, 2 M KOH or $(NH_4)_2S_2O_8$ solution was introduced into the above-mentioned solution at a rate of 2 mL·min$^{-1}$, until the pH value of the solution reached 11. After 30 min stirring, the milky solution was then transferred into a double-walled vessel, which has an inner liner and a cover made of Teflon perfluoroalkoxy (PFA) polymer resin and an outer high strength sleeve. The vessel was sealed and maintained in a Microwave Accelerated Reaction System (MARS-5, CEM Corporation, USA) at 140 °C for 3 h. The system was operated at 2.45 GHz with a maximum power of 1200 W. The system can be controlled by temperature (maximum, 300 °C) and pressure (maximum, 1500 psi). After the reaction was complete, the resultant black precipitates were centrifuged, washed with distilled water and then ethanol to minimize the extra ions in the final products, and dried at 80 °C in air. Finally, the powders were calcined at 270 °C to 500 °C in air for 2 h, respectively.

The thermal behavior of the precursor was examined using simultaneous differential scanning calorimetry (DSC) and thermogravimetric analysis (TGA, METTLER TOLEDO). The measurements were performed at a heating rate of 10 °C min$^{-1}$ from room temperature to 600 °C in air. The morphology and structure of the as-prepared samples were characterized by powder X-ray diffraction (XRD, GBC MMA, Cu Kα radiation, 40 kV, 25 mA), field emission scanning electron microscopy (FESEM, JEOL JSM7500FA, 15 kV), transmission electron microscopy (TEM), high-resolution TEM (HRTEM), and selected area electron diffraction (SAED) (JEOL 2011F, 200 kV). Specific surface areas and pore size distributions of the calcined samples were determined from the results of N2 physisorption at 77K (Micromeritics ASAP 2020) using the Brunauer–Emmet–Teller (BET) and Barrett–Joyner–Halenda (BJH) methods, respectively.

**Electrochemical measurements.**

The electrochemical measurements were carried out via CR2032 coin-type cells with lithium metal as the counter and reference electrode at room temperature. The working electrode consisted of 70 wt% active material (the as-prepared $Co_3O_4$), 15 wt% carbon black, and 15 wt% polyvinylidene difluoride (PVDF). The electrolyte was 1 M $LiPF_6$ in a 1:2 v/v mixture of ethylene carbonate (EC) and diethyl carbonate (DEC). Cells were assembled in an argon-filled glove box (Mbraun, Germany). Charge–discharge cycles were measured between 0.01 V and 3.0 V at a constant current density of 50 mA g$^{-1}$ on a Land CT2001A Cycler. Cyclic voltammetry (CV) of the cells was conducted at a scan rate of 0.1 mV s$^{-1}$ using an electrochemical workstation (CHI660C). The electrochemical impedance spectroscopy (EIS) was carried out on a PARSTAT 2273.

**Magnetic measurements.**

The magnetic measurements for the samples were carried out using the vibrating sample magnetometer (VSM) option of a Quantum Design 14 T Physical Properties Measurement System (PPMS) in the temperature range of 5 – 305 K at applied fields up to 5 T.

3. Results and Discussion
3.1 Microstructure observations and analysis

Different nanostructure morphologies (porous nanoplatelets, hexagonal nanorings, and porous nanowires) of $Co_3O_4$ were prepared by the calcination of a cobalt hydroxide precursor obtained by the large-scale synthesis of β-$Co(OH)_2$ nanoplatelets and nanowires via a microwave hydrothermal process, using potassium hydroxide (obtained nanoplatelets) and urea (obtained nanowires) as mineralizers at 110-140 °C for 3 h. Porous $Co_3O_4$ nanoplatelets (as shown in the field emission scanning electron microscope (FESEM) images in Supplementary Fig. S1(c) and (d)) were prepared by calcining the β-$Co(OH)_2$ nanoplatelets (FESEM images shown in Supplementary Fig. S1(a) and (b)) at 350 °C for 2 h (sample identified as P350), while the hexagonal $Co_3O_4$ nanorings, identified as P500 (FESEM images shown in Supplementary Fig. S1(e) and (f)), were prepared by calcining the β-$Co(OH)_2$ nanoplatelets at 500 °C for 2 h. For the transformation of β-$Co(OH)_2$ nanoplatelets into a series of distinct $Co_3O_4$ samples, the process of converting the β-$Co(OH)_2$ nanoplatelets into the porous $Co_3O_4$ nanoplatelets is a self-supported topotactic transformation, which is easily controlled by varying the annealing temperature [23]. When the temperature is raised to 500 °C, the morphology is transformed to hexagonal nanorings (as shown in Fig. 1(b)).

A second series of porous nanowires (FESEM images shown in Supplementary Fig. S2(c-j)) were prepared by calcining β-Co(OH)$_2$ nanowires (FESEM images shown in Supplementary Fig. S2(a) and (b)) at 300, 350, 420, and 500 °C for 2 h. The obtained Co$_3$O$_4$ samples are respectively defined as W300, W350, W420, and W500. For the β-Co(OH)$_2$ nanowires, subsequent calcination of this precursor at 300 °C, 350 °C, 420 °C, and 500 °C in air caused a spontaneous transformation of the morphology, forming Co$_3$O$_4$ porous nanowires with average diameters of 8, 15, 25, and 45 nm, respectively, and average lengths of few micrometers.

Figure 1 shows typical low-magnification transmission electron microscope (TEM) images of the synthesized porous nanoplatelets (P350) (a), nanorings (P500) (b), and porous nanowires W300 (c), and W350 (d). The inset in each figure shows the corresponding selected area electron diffraction (SAED) pattern. The SAED pattern reflects the degree of orientation of each sample. Clearly, the porous and textured Co$_3$O$_4$ nanostructures are *in-situ* transformed from the β-Co(OH)$_2$ nanostructures by the nucleation and growth of multiple Co$_3$O$_4$ crystals within the β-Co(OH)$_2$. The porous nature of the transformed material may be due to *in-situ* β-Co(OH)$_2$ decomposition, with the loss of H$_2$O. This further reveals that the nanostructures grow in an oriented manner to form an integrated porous architecture with interesting combined properties of porosity and quasi-single-crystallinity.

With pronounced thermal re-crystallization, the morphology of the Co$_3$O$_4$ calcined at 500°C for 2 h has changed significantly from the porous hexagonal platelets (calcined at 350°C) to hollow Co$_3$O$_4$ rings. The Co$_3$O$_4$ particles (40-50 nm) are also significantly coarser than those (10-20 nm) of the sample calcined at 350°C (Fig. 2(a)). Fig. 2(b) contains high resolution transmission electron microscope (HRTEM) images of an inner site and an outer site of the hollow ring as marked. The corresponding SAED patterns clearly reveal that the inner and outer surfaces of the ring have crystallized in different orientations.

We examined the crystallographic nature of the porous Co$_3$O$_4$ nanowires using HRTEM observations. Figure 2(c) shows part of a porous Co$_3$O$_4$ W300 nanowire. Except for a little coarseness in the particle size, there is not much difference in the HRTEM observations of W350, W420, and W500 nanowires. Figure 2(d) contains a section of a porous Co$_3$O$_4$ W350 nanowire. The insets show several nanowires and the individual nanowire from which the HRTEM image comes.

Taking all these TEM images into account, it is interesting to note that all the nanostructures (porous nanoplatelets, hexagonal nanorings, and porous nanowires) consist of layers of nanosheets with a few 0.47 nm (111) atomic plane fringes (white lines) and large amounts of 0.284 nm (220) atomic plane fringes (yellow lines), which directly reflect the exposed outer surface planes, which are indicated to dominate the properties of these transition metal oxide nanocrystals, especially the catalytic properties [5, 26]. These constituent nanosheets are single crystals 3-8 nm thick. This was confirmed by axial rolling TEM observations of an individual nanowire (images shown in Supplementary Fig. S3). Furthermore, the joint areas (the yellow circled areas in Fig. 2(a) and (d)) contain 0.47 nm (111) atomic plane fringes, which cross over 0.284 nm (220) atomic plane fringes, confirming the thin platelet crystallographic nature of the nanosheets, which are stacked to form porous nanostructures. Due to these crystallographic features of our nanostructures, it is difficult to make a precise quantitative determination of the average orientations of the nanosheets by HRTEM, the method employed by Xie et al. [5], but they can be determined qualitatively by SAED patterns, as shown in the insets of Fig. 1(a)-(d). We can quantitatively determine that the densities of the {110} plane textured crystals are in the order W350 > W300 > P350 > P500.

We have reported a facile comparative method to relatively precisely determine the average orientations of powder samples [27, 28], e.g., combining the results of powder x–ray diffraction with calculation results on the fully random powder diffraction state through the retrieval refinement Fullprof software package. For the details, see Supplementary Information Fig. S4. Fig. 3 shows the percentage of texture and the lattice parameters of porous nanowires prepared at different annealing temperatures. It shows that all the samples exhibit {110} plane texture, but sample W350 has the highest percentage of texture and the smallest lattice parameter. This means the $Co_3O_4$ nanosheets mainly grow along the [110] direction and preferentially expose the {110} plane. 350 °C is the optimum temperature compared with the other three temperatures to obtain the highest {110} plane surface area, which is estimated to be 64% of the total surface area from the texture data analysis.

The porous structure of the annealed $Co_3O_4$ samples can be quantified with nitrogen adsorption/desorption measurements. Fig. S6 shows the $N_2$ adsorption/desorption isotherms and corresponding Barrett–Joyner–Halenda (BJH) pore size distribution curves (insets) of selected samples (a) P350, (b) P500, (c) W350 and (d) W500. The characteristic type IV isotherms with type H3 hysteresis loops for both nanoplatelets and nanowires samples, which confirm the highly porous structure of the $Co_3O_4$ samples prepared by annealing the cobalt-basic-carbonate compounds. This result is consistent with the above FESEM, TEM and HRTEM observations. More quantitative measures come from the corresponding pore size distributions calculated from the desorption branches. As can be seen from the inset of Fig. S6, the pore size distribution is board, in the range of 9–50 nm, for nanoplatelets samples (Fig. S6 (a) and (b)), but it is very narrow, in the range of 9 –16 nm, for nanowire samples (Fig. S6 (c) and (d)). This indicates the inter-crystalline pores in the $Co_3O_4$ nanowires are well-proportioned and uniformity. The Brunauer–Emmett–Teller (BET) specific surface area is 34 $m^2$ $g^{-1}$ for P350 and 41$m^2$ $g^{-1}$ for W350.

## 3.2. Electrochemical and Li–ion battery properties

All the $Co_3O_4$ nanostructure samples were tested with respect to their electrochemical properties and evaluated for their Li–ion battery performance under the same conditions, e.g., the as-prepared $Co_3O_4$ products and lithium foil serve respectively as the positive and negative electrodes, and compose standard $Co_3O_4$/Li half cells. Fig. 4(a) and (b) shows the cycling performances of the electrodes made from the porous nanoplatelet series of samples, P270, P350, and P500, and the porous nanowire series of samples, W300, W350, W420, and W500, at a current density of 50 mA $g^{-1}$ (equivalent to 1 Li per formula unit in 2.2 h), which is commonly used in the energy materials community for testing $Co_3O_4$/Li, and a 400 mA $g^{-1}$ charge rate. For P270, P350, and P500, the first discharge capacities are around 1150 mA h $g^{-1}$. The initial coulombic efficiency of P350 was 81.9%, i.e. the porous $Co_3O_4$ nanoplatelets have an irreversible loss of only 18.1%, which is probably the lowest loss for $Co_3O_4$ reported so far. More importantly, for P350, after the initial decrease, the reversible capacity gradually increases to an ultra-high value of ~1540 mA h $g^{-1}$ over the first 42 cycles. After that, this capacity tends to be stabilized and maintained at about 1400 mA h $g^{-1}$ after the 100th cycle. The capacity of P350 is twice that of the large-sized multilayered $Co_3O_4$ platelets reported by Yao et al. [29], and it is even higher than those of reported $Co_3O_4$ nanotubes and nanowires [2, 30-32]. It is not unexpected that the cycling performance of P500 should be very different from that of P350, although their first discharge capacities are similar. The initial coulombic efficiency of P500 was only 66.1%, and the capacity rapidly faded to about 300 mA h $g^{-1}$ over the first 10 cycles. Similarly, for the porous nanowire series of samples (W300, W350, W420, and W500), the first discharge capacities at a current density of 50 mA $g^{-1}$ are nearly the same, with all above the theoretical capacity of $Co_3O_4$ of 890 mA h $g^{-1}$ and around 1000 mA h $g^{-1}$, but the initial coulombic efficiencies of the samples are very different, from 82% (W350) to 50%. Surprisingly, for W350, the reversible capacity gradually increases to an ultra-high value of ~1350 mA h $g^{-1}$ over the first 50 cycles, but it is reduced to ~790 mA h $g^{-1}$ over the next 50 cycles and to ~630 mA h $g^{-1}$ after 100 cycles at a current density of 400 mA $g^{-1}$. To the best of our knowledge, the performance of the $Co_3O_4$-based anode materials for lithium batteries presented here is the best to date [2, 33-36].

The significantly different Li–ion battery performance of different temperature treated nanoplatelets, nanoswires samples and between two series samples indicate that the electrochemical properties of $Co_3O_4$ are extremely sensitive to its structure, size and morphology. To understand the electrochemical process of these difference in behavior, series electrochemistry measurements on selected $Co_3O_4$ nanostructured electrodes made by both nanoplatelets and nanoswires samples were performed. Fig. S7 showed the cyclic voltammograms (CV) of electrodes made by nanoplatelets samples P350 and P500 for the first, second and the 5th cycle. In the first cycle, there are two reduction peaks including one weak shoulder peak at 1.2V and another strong peak at 0.9V and one oxidation peak at 2.1V for both P350 and P500. The two reduction peaks are generally attributed to the reduction of the $Co_3O_4$ to CoO (or $Li_xCo_3O_4$) and metallic Co, respectively, accompanying the oxidation of metallic Li. The oxidation peak is mainly attributed to the oxidation of metallic Co accompanying with the decomposition of $Li_2O$ [37]. The whole conversion reaction of Li with $Co_3O_4$ can be written as follows:

Discharging

$$Co_3O_4 + 8Li^+ + 8e^- \longleftrightarrow 4Li_2O + 3Co,$$

Recharging

which has a theoretical capacity of 890mAhg−1. For P350, the main reduction peak shifts to a higher potential at 1.1V in the subsequent cycles. This might be related to the pulverization of the $Co_3O_4$ [38]. A shoulder reduction peak at 0.6V and a shoulder oxidation peak at 2.4V appear in the subsequent cycles, indicating that another reversible redox reaction accompanies with the reaction between Li and $Co_3O_4$. This is attributed to the formation and decomposition of the polymer layer on the surface of the active material. The gradually increased capacity up to 1350mAhg$^{-1}$ over the first 30 cycles is likely to be due to the contribution of the polymer layer. The existence of higher than theoretical specific capacity in $Co_3O_4$ has also been observed and reported before [39, 40], and it is plausibly attributed to the reversible growth of a Li-bearing polymeric layer on the surface of the active material, resulting from electrolyte degradation. The above results show that the porous $Co_3O_4$ platelets have extremely high electrochemical activity, which is probably due to the nano- $Co_3O_4$ particles are rigidly confined within the hexagonal platelets and they are prevented from agglomeration and lithium ions are readily accessible from the electrolyte due to the mesopores within the hexagonal $Co_3O_4$ platelets. Both the reversible formation of $Li_2O$, which is known to be electrochemically inactive in bulk, and the reversible formation of the polymer layer catalyzed by cobalt or cobalt oxide nanoparticles can occur. It is well known that both nanosized metallic cobalt and cobalt oxide are excellent catalysts [39-42]. It should be noted that the shoulder peaks, corresponding to polymer layer formation and decomposition for P350, could not be observed in the CV curves for P500, furthermore, the reduction peak in the 5th cycle obviously shifts to a lower potential at 0.8 V, indicating an increase in the cell polarization that leads to the incompletely reversible redox reaction between Li and $Co_3O_4$, which causes the capacity to rapidly fade during cycling. The increase in the cell polarization for P500 is further confirmed by the electrochemical impedance spectrum (EIS) measured before and after 5 cycles on the same electrode (see Fig. S8 in Supplementary Information). The difference in electrochemical behavior of the porous $Co_3O_4$ nanoplatelets P350 and P500 is further explained by the discharge/charge profiles in Fig. S9. From the discharge profiles, it is clear that capacities are mainly from two sections, the long plateau region (corresponding to the redox conversion reaction between Li and $Co_3O_4$) and the sequent sloping region (corresponding to the surface layer formation). The first discharge curves for P350 and P500 are similar, due to undergoing the resemble initial conversion reaction, so their first discharge capacities are both approximate 1200 mAhg$^{-1}$.

However, the sequent cycles for P350 and P500 are significant different. For P500, 1st, 2nd and 30th discharge/charge curves show that the electrochemical reactions appear the poor reversible, for the 30th cycle, the plateau reaction almost ceases, and so that the discharge capacity is not more than 200mAhg$^{-1}$. For P350, 1st, 2nd, 30th and 60th discharge/charge curves indicate that the redox conversion reaction and the surface layer formation in the cycles are reversible, so the porous $Co_3O_4$ nanoplatelets exhibit an ultrahigh reversible capacity. In early cycles, more and more polymer layer formation makes the capacity gradually increase, for the 30th cycle the capacity is up to the maximum, thereafter the capacity gradually decreases due to little irreversible conversion, after the 60th cycle the capacity tends to be stable, so the capacity retention appears the semi-circle-type feature.

The difference in electrochemical behavior between the porous $Co_3O_4$ arrays self-assembled from porous nanowires W350 and W500 is further explained by the discharge/charge profiles in Fig. S10 (a) and (b). From the discharge profiles, it is clear that capacities are mainly from two sections, the long plateau region (corresponding to the redox conversion reaction between Li and $Co_3O_4$) and the subsequent sloping region (corresponding to the surface layer formation). Due to undergoing a similar initial conversion reaction, the early discharge/charge curves for W350 and W500 are similar, with their first discharge capacities both about 1200mAhg−1 and their initial coulombic efficiencies both close to 80%. However, the subsequent cycles for W350 and W500 are significantly different. For W350, the 40$^{th}$ and 60th discharge/charge curves indicate that the redox conversion reaction and the surface layer formation/decomposition with cycling are reversible, so the porous $Co_3O_4$ nanowires exhibit an ultra-high reversible capacity. In the early cycles, more and more polymer layer formationmakesthe capacity gradually increase, and for the 40th cycle, the capacity is up to the maximum. Thereafter, the capacity gradually decreases because there is little irreversible conversion, and after the 60th cycle, the capacity tends to be stable, so that the capacity retention appears as a semicircle-type feature. We propose that the higher surface area of the porous nanowires permits, on cycling, an increase in the extent of polymer layer formation on the surface of the active materials. For W500, the 15th and 30th discharge/charge curves show that the electrochemical reactions appear to be poorly reversible. For the 30th cycle, the plateau reaction is obviously shortened, so that the discharge capacity is not more than 400mAhg−1.

To further elucidate the electrochemical process, cyclic voltammograms (CV) of sample W350 for the first, second, and 5th cycles are presented in Fig. S8 in the ESI. In the first cycle, there are two reduction peaks, including one weak shoulder peak at 1.2V and another strong peak at 0.9 V, and one oxidation peak at 2.1 V. The two reduction peaks are generally attributed to the reduction of the $Co_3O_4$ to CoO (or $Li_xCo_3O_4$) and metallic Co, respectively, accompanying the oxidation of metallic Li. The oxidation peak is mainly attributed to the oxidation of metallic Co, accompanying the decomposition of Li2O [36]. The whole conversion reaction of Li with $Co_3O_4$ has a theoretical capacity of 890mAhg−1. The main reduction peak shifts to a higher potential at 1.15V in the subsequent cycles. This might be related to the pulverization of the $Co_3O_4$ [37]. A shoulder reduction peak at 0.6V and a shoulder oxidation peak at 2.4V appear in the subsequent cycles, indicating that another reversible redox reaction accompanies the reaction between Li and $Co_3O_4$. This is attributed to the formation and decomposition of the polymer layer on the surface of the active material. The gradually increased capacity up to 1520mAhg$^{-1}$ over the first 30 cycles is likely to be due to the contribution of the polymer layer. The existence of higher than theoretical specific capacity in $Co_3O_4$ has also been observed and reported before [43-45], and it is plausibly attributed to the reversible growth of a Li-bearing polymeric layer on the surface of the active material, resulting from electrolyte degradation. The polymer layers that have been intensively studied turn out to be formed by a multistep degradation mechanism including electrochemical and chemical processes [46].

The Li-ion battery performance of W350 is quite promising in terms of capacity, cycling, and charge rate performance when compared with the $Co_3O_4$ previously reported [2, 33-36, 47-49], and even more interesting is its mechanism, as the capacity in samples W350, P270, and P350 is nearly double the theoretical capacity. In order to deeply understand the relationships between the electronic structure, the crystallographic structure, and the properties of nanostructures, to further derive the mechanism, we performed X-ray photoelectron spectroscopy (XPS) and investigated the magnetic properties of the $Co_3O_4$ nanostructures, since they are very sensitive to the ion valence state and electron state of the nanostructures. Another surprise was that sample W350, which had the best Li-ion battery performance, presented unusual room temperature ferromagnetism in this conventionally antiferromagnetic material, which motivated us towards a further detailed investigation of these $Co_3O_4$ nanostructures.

### 3.3 Abnormal magnetic properties and room temperature ferromagnetism (RTFM).

The magnetic measurement results are shown in Fig. 5, in which the magnetisation, (with zero-field cooling (ZFC) and 1 kOe field cooling (FC)) versus temperature (M – T) curves are shown in (a) nanoplatlets and (b) nanowires, the magnetisation versus applied field (M – H) hysteresis loops are shown in c) nanoplatlets and (d) nanowires. In bulk $Co_3O_4$, the magnetic transition from the paramagnetic state to antiferromagnetic occurs at $T_N$ = 42 K. It should be noted that the Néel temperature is smaller than the bulk value of 42 K due to the well-known finite-size effect [50-52], and not to dilution effects in the core. The M vs T curves measured for both $Co_3O_4$ nanoplatlets and nanowires series samples show a splitting (bifurcation) of the FC and ZFC magnetization below a temperature $T_{bf}$. The results of nanowires samples are simply reported in our previous studies in ref. [28, 48, 49], and are agreement with previous studies on $Co_3O_4$ nanowires reported by Ref.33. It is well known that AF nanosystems are usually governed by core-shell behavior. The wire cores show regular AF order, whereas the surface is the most sensitive part of the materials to any kind of treatment or interaction. Thus, nominally identical oxides nanostructures can present different magnetic properties due to different surface states. This could account for the discrepancies in results from different groups that use nanoparticles without well-defined surfaces state to claim the different origin, and findings from different groups are commonly contradictory. So that, precise identification of the nature of the surface contribution has remained unclear. Terms such as ''disordered surface state,'' ''loose surface spins,'' ''uncoupled spins,'' "diluted AFM", ''spin-glass-like behavior,'' "high anisotropy energy step-edge orbital moment," etc. express the uncertainty in the description of the surface contribution. Several experimental studies followed, suggesting various scenarios for the magnetic properties found, e.g., surface roughness and surface step atoms make remarkably different contributions to magnetic moment, spin-glass or cluster-glass-like behavior of the surface spins [53-56], thermal excitation of spin-precession modes [57], finite-size induced multi-sublattice ordering [58, 59], core-shell interactions (diluted AFM shell) [33-36], or weak ferromagnetism [60, 61]. This report will compared different magnetic behaviours between nanoplatelets and nanowires, focus on the exploration of the origan of room temperature ferromagnetic behaviour (RTFM) occurring in W350 sample. Our results and detail analysis indicate that, it is more reasonable for using the core-shell structure and core / shell magnetic interaction scenario to interpret the abnormal magnetic behavours and RTFM emerging in the samples.

In Fig.5(b), the M-H loops measured at 5K below the $T_N$, loops should be shown straight lines, e.g. AFM features, most of them does at low field range (<1T shown in the figure). However, the W350 sample's loop shows an typical ferromagnetic features, and M-H hysteresis loop of P270 sample show very week FM features as well. The M-H hysteresis loops were further charactered at room temperature ranges, as shown in the insert in Fig. 5(b), surprisily it shows the same features as 5K for both W350 and P270 samples. The M-H hysteresis loops for both W350 and P270 show far away saturated feature e.g. show the combination of two components. As discussed above, the core should not change much, AFM core, but magnetic properties of the shell layers show variation, they show RTFM for W350 and P270. So the M-H hysteresis loops are combined core and shell magnetic behavours with additional core-shell magnetic interact each other. At high temperature range (> $T_N$), the M-H hysteresis loops are the combination of paramagnetic core with ferromagnetic shells for W350 and P270, and with other magnetic behaviors shell for other samples. At low temperature range (< $T_N$), the M-H hysteresis loops are the combination of antiferromagnetic core with ferromagnetic shells for W350 and P270, and with other magnetic behaviours (such as shell for other samples. In addition, at $T < T_N$, the interaction of AFM core with different magnetic behaviours shell would affect magnetic behaviours of the samples, e.g. exchange interaction of AFM/FM, AFM/DAFM (diluted AFM), or AFM/Spin-glass-like, etc. the M-H hysteresis loop may display an enhancement of the coercive field ($H_C$), and an vertical or horizontal shift due to the interfacial coupling, e.g. exchange bias ($H_{EB}$) effects when field cooling the samples lower than $T_N$. Although the displayed shifts may include two contributions: (i) unsaturation remanence magnetisation when the applied filed lower than the so-called reversal field [62, 63] (ii) the real exchange bias ($H_{EB}$). In order to consistent with the exchange bias research community, we here donate the horizontal shift as $H_{HS}$, since we didn't determine the reversal field yet. We have systematically measured magnetization hysteresis loops at 5 K after ZFC and FC on $Co_3O_4$ nanostructures, the summary results, are shown in Fig. 6. Selected three of magnetization hysteresis loops are shown in Fig. S12, (a) the hysteresis curve measured after FC in 1 kOe displays an enhancement of the coercive field from $H_C$ = 7Oe (ZFC) to 24 Oe (FC) and the horizontal shift $H_{HS}$ = -94Oe (ZFC) to -168Oe (FC) values for P270 sample; (b) the hysteresis curve measured after FC in 50 kOe displays an enhancement of the coercive field from $H_C$ = 63Oe (ZFC) to 74 Oe (FC) and the horizontal shift $H_{HS}$ = 7Oe (ZFC) to -130Oe (FC) values for W350 sample; (c) the hysteresis curve measured after FC in 50 kOe displays an enhancement of the horizontal shift $H_{HS}$ = -30Oe (ZFC) to -600Oe (FC) values, but the coercive field do not display any enhancement, $H_C$ = 20Oe for both ZFC and FC for W500 sample.

Fig. 6 presents the $T_N$, the coercive field ($H_C$), the horizontal shift ($H_{HS}$) under 1 kOe and 50 kOe FC, and the average grain size (for easy comparison) versus annealing temperature. From these results, we can conclude that (i) $T_N$ is consistent with previous reports [64-66] of a decrease with increasing annealing temperature, due to the decreasing grain size; (ii) the $H_{HS}$ behaviours of most of the samples are as the same as in Refs: [67, 68], consistent with an antiferromagnetic (AFM) $Co_3O_4$ core and diluted AFM or spin glass – like shell, except for W350. This sample's $H_{HS}$ behaviour presents unusual features, which may be because the magnetic state of the shell is different, e.g. the magnetic state of the shell of W350 nanowires changes to ferromagnetic behaviour from normal diluted AFM or spin glass – like behaviour; (iii) sample W350 presents obvious room temperature, even high temperature, ferromagnetism, and sample P270 presents very weak room temperature ferromagnetism (as shown in the inset of Fig. 5(b)).

**3.4 Evaluations of conductivity by magnetic susceptibility, magnetisation hysteresis loops and electrochemical impedance dynamics measurements.**

It is well known that the magnetic properties of nanostructures are strongly affected by the surface to volume ratio or particle or grain sizes, when the size is reduced an increasing fraction of atoms lies at or near the surface and then surface and interface effects become more and more important. The presence of defects, broken exchange bonds, fluctuations in the number of atomic neighbours, and interatomic distances induce surface spin disorder and frustration. The surface magnetic behaviours displayed different features due to the variation of the surface structure, morphology, defects concentration (most likely oxygen vacancies), especially the expose face or orientation as specially deal with in this paper. The electron state at the surface layer may present delocalised free charge carries, they may week or strong interacted each other which may reconstructed the surface electron and crystal structures. Such that the surface magnetic state may present free electron / charge carriers Pauli-like paramagnetism, week or strong interacted spin-glass-like, or localised polarized ferromagnetic behaviours. These will significant affect the charge transport properties as well. For example, the free electron present in this semiconductor samples would significant enhance the electric conductivity, which would disaster improve the performance of electrochemistry cells.

Here we will continue to rich the technique that, by combining the magnetic susceptibility and magnetisation hysteresis loops measurement and fitting analysis to evaluate the charge transport properties, in these $Co_3O_4$ unique nanostructures. Since the magnetic susceptibility reflect free electron and unpaired electron features, and it is the average signal from few ten grams powder, the magnetic susceptibility measurement can be used to characterize the electron/charge transport in the large scale nanomaterials. The magnetic susceptibility measurement have been successfully used in investigating the charge transfer behaviour in lithium ion batteries previously [23-25, 47-49].

The temperature dependence of the magnetic susceptibility ($\chi$) in a field $\mu_0H$ = 1KOe for all samples are plotted in Fig. 5(a). The temperature dependences of $\chi$ and the Curie-Wiess constant, $C_{CW}(T) \equiv \chi T$, for samples are selected shown in Fig. 7(a). All $\chi T$–$T$ curves exist a partly linearly $T$-dependent Pauli-like susceptibility. The $\chi$–$T$ curves data (here we selected the data of 40 < T < 300K temperature range) can be fitted quite well using the sum of a Curie-like susceptibility, $\chi_C$, and a linearly $T$-dependent Pauli-like susceptibility, $\chi_P$:

$$\chi = \chi_C + \chi_P = [C_{CW} / (T + \Theta_{CW})] + [\chi_0 + \chi_1 T] \qquad (1)$$

Where, for example, $\chi_0$ = 1.23 × 10$^{-5}$ *emu / mol*, $\chi_1$ = – 9.96× 10$^{-7}$ *emu / mol K*, the *Curie-Weiss* temperature, $\Theta_{CW}$ = -217.6 K, and the *Curie* constant, $C_{CW}$ = 2.73 *emu / mol*, e.g. $\mu_{eff}$ = 4.67 $\mu_B$ for P270, and $\chi_0$ = 2.4 × 10$^{-4}$ *emu / mol*, $\chi_1$ = – 4.6 × 10$^{-8}$ *emu / mol K*, the *Curie-Weiss* temperature, $\Theta_{CW}$ = −83.6 K, and the *Curie-Wiess* constant, $C_{CW}$ = 5.4 × 10$^{-4}$ *emu / mol*, e.g. $\mu_{eff}$ = 0.14 $\mu_B$ for W350. The fitting line is shown by the blue dash line in Fig. 6(A). The fitting and calculation results for all samples are plotted in Fig 6(B), these results indicate that all samples exist conduction "free electrons/carriers" Pauli-like susceptibility ($\chi_0$), however the value of $\chi_0$ is much different in different sample. Comparing the Pauli-like susceptibility $\chi_0$, it can be seen that the highest value existed in W350 sample with nanotubes of $\chi_0$ = 2.4 × 10$^{-4}$ *emu / mol*. It is observed that the value of $\chi_0$ in W350 is much larger than that in others. Selected magnetization *(M-H)* loops at temperature 305 K and under field up to 1 T are plotted in Fig. 6 (C) for both P270 and Fig. 6 (D) for W350. Following the suggestion from Eq. (1) that the system can be described in terms of the sum of contributions from *Pauli* and *Curie-like* terms, we fit the magnetization experimental data to:

$$M(H) = \chi^* \times H + M_0 B \{S, H, T\} \qquad (2)$$

$B\{S, H, T\}$ is the Brillouin function, it can be fitted with the dominant contribution $\chi^*$ = 1.6 × 10$^{-5}$ *emu mol$^{-1}$ Oe$^{-1}$* and $M_0$ = 3 × 10$^{-4}$ $\mu_B$ / f.u. for P270 sample, and $\chi^*$ = 1.01 × 10$^{-4}$ *emu mol$^{-1}$ Oe$^{-1}$* and $M_0$ = 2 × 10$^{-4}$ $\mu_B$ / f.u. for W350 sample. The fitting results and disperse the two parts curves are plots in insert of Fig.6 (C) and (D). In P270, the behaviour of $B\{S, H, T\}$ shows very weak ferromagnetism with coercivity field $\mu_0H_C$ = 78 Oe. The value of $\chi^*$ corresponds 1.6 × 10$^{-5}$ *emu mol$^{-1}$ Oe$^{-1}$* and $M_0$ corresponds 3 × 10$^{-4}$ $\mu_B$ /f.u., suggesting only a small number of "free-electrons" contribute to the first term and a large number of spins contribute to the second term in Eq. (2). In W350, the behaviour of $B\{S, H, T\}$ is typical of a ferromagnetism with coercivity field $\mu_0H_C$ = 126Oe. The value of $\chi^*$ corresponds 1.01 × 10$^{-4}$ *emu mol$^{-1}$ Oe$^{-1}$* and $M_0$ corresponds 2.3 × 10$^{-4}$ $\mu_B$ /f.u., suggesting a large number of "free-electrons" contribute to the first term and only a small number of spins contribute to the second term in Eq. (2). Above fittings and analysis indicated that the W350 sample presents 1.01 × 10$^{-4}$ / 1.6 × 10$^{-5}$ > 6 time "free-electrons or charge carries" higher than the P270 sample at room temperature, the magnetic susceptibility measurements (Fig.6) confirm the "free-electrons" features, which suggest that the density of freedom electrons, which related to the electric conductivity in W350.sample should much higher than that in P270 sample. There are not ferromagnetism present in other samples, so "free – electron or charge carries" components in susceptibility-$\chi_0$ should direct response to the conductivity. For comparison, the susceptibility fitting results vs annealing temperatures for two serial samples are plotted in Fig. 6(B). From the analysing of results shown in Fig.6, the $\chi_0$ values of two serials samples display that

$$W350 > P270 > W300 > P350 > W\ 420 > P420 > W\ 500 > P500,$$

the electronic conductivities of two serials samples should following the same tendance as the $\chi_0$ values, since the "free – electron or charge carries" in the samples may dominant the electronic conductivities of the samples.

In order to confirm the conductivity of samples evaluated by the magnetic susceptibility measurements, we have conducted a comparative investigation of the electrochemical dynamics of the selected two samples (P350 and W350). Impedance measurements were carried out on Li-ion batteries containing electrodes made from the two samples. These batteries were discharged to 0.6 V at different temperatures, and the results are shown in Support information Fig. S13 (a) and (b). The samples conductivities, i, fit the Arrhenius equation,

$$i(T) = i_0(A) = i_0 \exp(-\ E_a\ /\ kT),$$

where $i_0$ is the pre-exponential factor, $E_a$ the activation energy, and k Boltzmann's constant, as shown in Fig. 8. It can be obtained that W350 exhibits a lower $E_a$ = 44.5 kJ/mol than P350, where $E_a$ = 51.8 kJ/mol. Comparing the lattice parameter, a, in each sample, W350 also passes through a minimum value a = 0.8084 nm, as shown in Fig. 3, which leads to the conclusion that the activation energy $E_a$ increases with the lattice parameter and hence with the distance for electron hopping. These indicate that the W350 sample may possess higher conductivity as well.

### 3.5 XPS results, surface compositions and electron structure

$Co_3O_4$ has a spinel structure containing $Co^{3+}$ in an octahedral coordination (***O***) and $Co^{2+}$ in a tetrahedral coordination (***T***). Oxygen anions form a distorted face-centred cubic sublatttice, in which $Co^{2+}$ cations occupy one-eighth of the tetrahedral interstices and $Co^{3+}$ cations occupy half of the octahedral interstices. The HRTEM observations and XRD orientation calculations indicate that the surface mainly presents {111} and {110} planes. Clearly, the {111} planes contain only $Co^{2+}$ cations, while the {110} planes are composed mainly of $Co^{3+}$ cations. In fact, surface differential diffraction studies have proved that the $Co^{3+}$ cations are present solely on the {110} planes [69, 70]. Therefore, for our samples, especially W350, it is most likely that $Co^{3+}$ cations on the exposed surface planes in the ***O*** position of $Co_3O_4$ nanostructures are electron trapped $Co^{3+}|\bullet$ state or reduced to $Co^{2+}$ cations most likely is due to the surface oxygen deficiency, defects or abnormal surface energy states stabilised special {110} orientation state.

In order to identify and to confirm these hypothesizes, the subsequent XPS results further confirmed this hypothesis, and in addition, several groups [71-73] have found room temperature ferromagnetism in $Co_3O_4$ and identified the original microscopic mechanism as due to $Co^{2+}$ ions replacing $Co^{3+}$ in the ***O*** positions by different methods. In our case, the XPS results on the $Co_3O_4$ nanostructures are shown in Fig. 9, in which the Co characteristic peaks Co 2p1/2 and Co 2p3/2 are shown in (a), carbon characteristic peaks (C 1s) are shown in (b), the oxygen characteristic peak O 1s is shown in (c), the relative surface concentrations of carbon and oxygen to cobalt estimated using integrated XPS intensity values ($I_C/I_{Co}$ and $I_O/I_{Co}$) are shown in (d), and the whole range of peaks is shown in supporting information Fig. S14. The as-prepared samples may absorb several surface contaminants, the most common being potassium, carbon, and calcium. These contaminants are estimated to be in submonoloayer coverage based on their Auger intensities. These compositional information can be obtained from XPS data. From the XPS results, we can conclude that all samples are high purity, except very small amount of carbon, which can rule out other impurity and metallic Co, Fe etc. contaminating magnetic particle effects. For the relative surface concentrations of carbon and oxygen to cobalt can be estimated using integrated XPS intensity values ($I_i$) and the appropriate sensitivity factors ($S_i$) with $S_{Co}/S_C \sim S_{Co}/S_O \sim 5$.[74, 75] The calculated concentration ratios of carbon and oxygen to cobalt ($C_C/C_{Co}$ and $C_O/C_{Co}$),

$$\frac{C_C}{C_{Co}} = \frac{I_C/S_C}{I_{Co}/S_{Co}} \text{ and } \frac{C_O}{C_{Co}} = \frac{I_O/S_O}{I_{Co}/S_{Co}},$$

are plotted in Fig. 9(d), the carbon concentrations for all samples are around 0.3, no special cases; the oxygen concentrations around 1.288 are lower than the theory stoichiometric ratio 1.33 within error, indicate that all samples present different degrees of oxygen vacancy. In addition, it is interesting to note that carbon contaminates and the oxygen concentrations in the samples are strong dependent on the preparation conditions, the tendencies are decreased with the annealing temperature increasing for both nanoplatelets and nanowires samples; in additional, the absorbed carbon contaminants and oxygen contents in nanoplatelets are higher than that in nanowires samples, that indicated the effects of preparation condition, seems that using urea (obtained nanowires) as mineralizers can obtain much more purity and more completely $Co_3O_4$ crystalline that of potassium hydroxide (obtained nanoplatelets).

A comparison of cobalt 2p regions for selected $Co_3O_4$ nanostrucutres is shown in Fig. 9 (a). $Co_3O_4$ in its various nanostructures of preparation and $Co^{3+}$-containing have main binding energy peaks at 779.8 and 794.8 eV for 2p3/2 and 2p1/2 (the blue lines in Fig. 9(a)), respectively. It also can be noted that there are higher binding energy splitting out peaks at 780.9 and 796.2 eV for 2p3/2 and 2p1/2 (the red lines in Fig. 9(a)), respectively, those are $Co^{2+}$-containing 2p3/2 and 2p1/2 peaks in $Co_3O_4$ 2p binding energies. We noted that $Co^{2+}$-containing in CoO 2p binding energies are do higher, 780.5 and 796.6 eV in Ref.[76-79]. Weak 2p satellite features for the spinels (two marked as blue arrows and S in Fig.9 (a)) are found at 790.8 and 805.2 eV, with significantly reduced intensity compared to the intense CoO satellites[76-81], which are found at lower binding energies.[77, 79] The weak satellite structures of $Co_3O_4$ are characteristic of spinel structures in which $Co^{3+}$ cations occupy octahedral lattice sites with diamagnetic, filled $t_{2g}$ and empty $E_g$ levels, and $Co^{2+}$ cations are in tetrahedral sites.[78-81] $Co^{2+}$ cations in CoO are high spin $d^8$ in octahedral lattice sites and the intense satellite structure has been proposed to result from the charge-transfer band structure found in late 3d transition metal monoxides with partially filled $E_g$ character.[78, 80, 82]. These weak satellite structures of $Co_3O_4$ are supported by studying in the surface composition and structure of single crystal $Co_3O_4$ (110)[83,]

The unpaired nature of the half-filled $E_g$ band of the $Co^{2+}$ cation in CoO results in strong electron correlation and substantial broadening of the cobalt 2p main peaks due to many closely lying 3d final states in photoemission. In contrast, the low-spin, diamagnetic nature of the $Co^{3+}$ octahedral cation and the weaker crystal field effect of tetrahedral coordination for the $Co^{2+}$ cation result in similar photoemission binding energies and a sharper 2p peak for the spinel, despite the existence of two different cobalt oxidation states in this material. Thus the satellite structures, not absolute 2p3/2 and 2p1/2 binding energies, are better able to distinguish between rocksalt CoO and cobalt-containing spinels with octahedrally coordinated $Co^{3+}$. These indicted that the satellite structures are characteristic of a CoO-like environment in the {110} surface. Since the oxygen to cobalt atomic concentration are around 1.288 (1.284 for W350 sample) as measured by XPS above, it still present oxygen vacancy at surface, which make the CoO-like environment in the $Co_3O_4$ (110) surface.

The O 1s region for selected $Co_3O_4$ nanostructures samples shown in Fig. 9(c). The main oxygen peak due to lattice $O^{2-}$ is set to 529.9 eV, as has been previously found for CoO,[79] $Co_3O_4$[76, 79, 80, 82-84]. There are three satellite peaks at higher binding energy, 531.3, 532.4, 533.6 eV with much lower intensity than the main peak in W350 sample (shown in insert of Fig. 9(c)). The O 1s peak at 531.3 eV is comparable to that reported for surface carboxylate [84], surface hydroxyls,[76, 81] under-coordinated lattice oxygens (O⁻),[85, 86] chemisorbed oxygen,[42] and inaccuracies in the peak fitting due to the inability to reproduce the exact peak shape and/or secondary electron background. O 1s peaks with comparable binding energy have also been observed on $Co_3O_4$ thin films,[79, 80] on $Co_3O_4$ powder surfaces,[84, 87] and in a $Co_3O_4$ single crystal cleaned in UHV [77]. While it is not possible to rule out low levels of hydroxylation etc. absorbents, the O 1s peak at 531.3 eV satellite peak is probably at least partially attributable to the intrinsic O 1s peak structure, which is imperfectly reproduced in the peak fitting procedure. One indication of this in the present set of spectra is the broadness of the peak (full width at half maximum (FWHM) = 2.1 eV] compared to the lattice peak (FWHM = 1.2 eV). The very weak feature observed at 532.4, 533.6 eV in Fig. 9(c) may be attributed to surface defects (oxygen vacancies) or the lower binding energy shoulder of the Co Auger (L2M23V) transition.[40] Comparing the samples listed in Fig. 9 (c), it clearly shows that the W350 sample present the highest intensity of O 1s peak at 531.3 eV, which indicate the surface of W350 sample have more $Co^{2+}$ cations, or more surface defects (oxygen vacancies), or surface hydroxyls, or under-coordinated lattice oxygen (O⁻) e.g. $Co^{3+}|\bullet$, chemisorbed oxygen etc. These results and analyses consistent with the XRD and TEM observation of more (110) faces expose to the surface in W350 sample. From the XPS results, we can conclude that (i) the W350 sample is high purity, which can rule out impurity and metallic Co magnetic particle effects; and (ii) the W350 presents a higher amount of $Co^{2+}$ or $Co^{3+}|\bullet$.

In the $Co_3O_4$ composition, the weak superexchange interaction $Co^{2+}- O^{2-}- Co^{3+}- O^{2-}- Co^{2+}$ maintains the antiferromagnetism [88, 89]. The predominant interaction is the AFM superexchange between the **T** and **O** positions, while interactions **T - T** and **O - O** are also AFM, but much weaker than the **T - O** one. Consequently, there are two magnetic sublattices corresponding to **T** and **O** positions with antiparallel orientation. $Co^{3+}$ ions in **O** positions present no magnetic moment due to the large splitting of the 3d orbital in this symmetry [90]. Hence, in $Co_3O_4$, only $Co^{2+}$ in the **T** position holds a magnetic moment, and the **T - T** weak AFM interaction is the dominant one, so the system presents AFM behavior with a much lower ordering temperature of $T_N$ = 42 K. However, for $Co^{2+}$ atoms in an octahedral field, the orbital splitting is quite small, and a $Co^{2+}$ ion in this symmetry should hold a magnetic moment. For a region of the crystal where $Co^{3+}$ ions in **O** positions are electron trapped $Co^{3+}|\bullet$ state or reduced to $Co^{2+}$ cations, there should be magnetic moments in both **O** and **T** positions, with partially filled $t_{2g}$ orbitals. In this situation, the **T - O** AFM interaction would be the dominant one, and the system should present a behavior similar to that of $Fe_3O_4$, that is, ferrimagnetism (FIM) with a high ordering temperature due to the strong interaction of **T - O**. Consequently, the observed weak FM signal can be explained by the reduction of $Co^{3+}$ in **O** positions to $Co^{2+}$, as the XPS results demonstrate. However, the effect in $Co_3O_4$ will be restricted to very small surface regions; the spinel structure of $Co_3O_4$ is unstable if a large fraction of $Co^{3+}$ is reduced to $Co^{2+}$, promoting the transformation to CoO, which is AFM. Thus, it is not possible to obtain a bulk FM or FIM material, nor a bulk material with uniform magnetic properties based on this effect. Fe, Mn, Ni, etc. magnetic element doping is another story [91-94], but ferromagnetism certainly is easy to obtain at the surface. The surface is the most sensitive part of a material to any kind of treatment or interaction. Properties depending on the surface and its reactivity in the first stages can be completely different, as we found for the two sets of samples analyzed here, such as the porous nanoplatelets (P350 and P500) and nanowires (W300, W420 and W500), which do not present room-temperature ferromagnetism (RTFM), while only W350 presents RTFM. (Sample P270 presents negligibly weak RTFM.) This RTFM phenomenon in W350 could arise from the greater amount of {110} planes exposed at the surface, which allows more $Co^{3+}$ exposed at the surface to be reduced to $Co^{2+}$ (since the surface easily loses oxygen, causing oxygen vacancy, etc. defects), and this breaks down the balance between the antiparallel magnetization in the $Co^{2+}$ sublattices because the $Co^{3+}$ is reduced to $Co^{2+}$ at the surface, causing the surface RTFM. Our magnetic measurements support the surface RTFM phenomenon: the main contribution is the bulk AFM, while the weak FM is added to it and its effects are superimposed (as shown in Fig 5(a)). The magnetism may be too weak to be detected in other samples, but we have detected a very weak RTFM signal from the P270 sample (as shown in the inset of Fig. 5(b)). This also means that processing to transform β-Co(OH)$_2$ nanosheets and nanowires to $Co_3O_4$ nanostructures might not achieve optimum conditions. We have performed thermogravimetry (TG) and differential thermal analysis (DTA), and the results (Supplementary Information Figure S5) show the different decomposition temperatures for the nanosheets and nanowires, which can act as guidance for the optimum calcining temperature (in this case 270 ºC for nanosheets and 350 ºC for nanowires as approaching the optimized temperatures), period, etc. parameters, although we have still not reached the optimum conditions to achieve room temperature ferromagnetism.

According to the results and analysis above, we have plotted a schematic diagram to easily interpret the mechanism of surface RTFM in the pure $Co_3O_4$ nanomaterials. Fig. 10 shows the crystal structure of a $Co_3O_4$ sample with only Co ions shown (a); part of the cell lattice structure with $O^{2-}$ shown (b), and after oxygen is lost (c); and the surface atomic configurations in the {111} (d), {110} (e), and {110} (f) planes after oxygen vacancy has caused surface $Co^{3+}$ to alter to $Co^{2+}$. In the $Co^{3+}$ array as shown in the in (f) part of the $Co^{3+}$ has changed to $Co^{2+}$, which causes the surface antiferromagnetic configurations to become inhomogeneous, changing to an incommensurate AFM order, so as to generate surface room and even high temperature ferromagnetism. We have explained the origin of surface RTFM above in the exposure of {110} planes and their $Co^{3+}$ ions at the surface, because the easy loss of oxygen at the surface causes oxygen vacancies and changes the $Co^{3+}$ in an octahedral position (***O***) to $Co^{2+}$, which breaks down the balance in the antiparallel AFM configuration, creating surface room temperature ferromagnetism. Since the alteration of surface electronic structure and morphology might change the surface electronic and ion transport properties, the Li-ion battery performance might also change. We have performed indirect observations of the electronic conductivity by observing the electron charging in SEM and FESEM, where the observation results show that the electronic conductivity of the above-mentioned samples increases with increasing {110} plane orientation ratios.

**3.4 Relationships between microstructure, electrochemistry / Li-battery and magnetic properties – the electronic /charge transport properties evaluation.**

We now explore the mechanism of Li reactivity and the superior Li-ion battery performance in our samples. Apart from the classical Li insertion/deinsertion or Li-alloying processes, Poizot et al. [92] pointed out that the mechanism of Li reactivity involves the formation and decomposition of $Li_2O$, which accompanies the reduction and oxidation of metal nanoparticles (in the range of 1-5 nanometres), respectively, in transition-metal nanoparticles as negative electrode in the Li-ion battery. They expect that the use of transition-metal nanoparticles to enhance surface electrochemical reactivity will lead to further improvements in the performance of lithium-ion batteries. Based on this mechanism, it can be obtained that based on $Co_3O_4$ + 8 $Li^+$ + 8$e^-$ → 4 $Li_2O$ + 3 Co, the theoretical discharge capacity of $Co_3O_4$ should be 890 mAh/g, as mentioned above. However, the highest discharge capacity among our porous nanowire samples, for W350, exceeds the theoretical value with 1540 mAh/g at the 50 mAh/g rate, which is nearly double the theoretical capacity, so it seems that this phenomenon cannot be interpreted by only using one mechanism for the formation and decomposition of $Li_2O$. This phenomenon of exceeding the theoretical value has been observed by other groups [32, 94], and they interpreted it as due to lithium storage in a polymer layer that forms on the high surface area of the nanowire [32, 47-49], which is facilitated by the porous structure and small size of $Co_3O_4$ [94]. Our results seem to support the following mechanism: higher lithium storage capacity may result from the intercalation of Li into

$Co_3O_4$ up to x ≈ 1.5 Li in $Li_xCo_3O_4$, so that both materials form a nanocomposite of Co particles embedded in $Li_2O$, which on subsequent charge forms CoO. The increase in the capacity on cycling from the initial cycles to values exceeding the theoretical value is interpreted as due to lithium storage in a polymer layer that forms on the high surface area of porous nanostructures of $Co_3O_4$ [32, 47-49].

However, our Li-ion battery results on the P270 and W300 porous sample with smaller grain and particle size and higher surface area is contrary to the explanation depending on the high surface area of porous nanostructures. Here, we attribute the higher Li-ion battery performance to the higher {110} plane orientation and porous nanostructure for the following reasons: **(i)** The higher the {110} plane orientation with more $Co^{3+}$ exposed on the surface of the nanostructured grains present, the higher the strength of catalytic effects at the surfaces of the nanostructures will be on the reaction of Li with $Co_3O_4$, which promotes the formation of a $Li_xCo_3O_4$ polymer layer on the surface and active carbon layers, even graphene layers, that all contribute to the Li capacity. The formation of $Li_xCo_3O_4$ polymer layers has been confirmed by other groups [2, 32], and our TEM and HRTEM observations on the samples after electrochemical cycling further confirmed the formation of polymer layers, as well as the formation of active carbon layers, even graphene layers, as shown in Fig. 11. Actually, strong supporting information is provided by the observation of extremely strong CO catalytic effects on nanorods with {110} planes exposed [5]. These active carbon or graphene layers, which are formed during electrochemical cycling, further improve the electronic conductivity of the electrode. **(ii)** The higher the {110} plane orientation the nanostructured grains present, the higher the electronic conductivity of the surfaces of the nanostructures, which has been confirmed by electron structure analysis through magnetic properties and electronic conductance in the SEM and FESEM analyses above, and also by the better battery cycling performance. **(iii)** The porous morphology of the nanostructures promotes Li ion transport, as they can accommodate large strain without pulverization, provide good electronic contact and conduction, and provide short lithium insertion distances. **(iv)** In addition, another possible mechanism is that the higher the {110} plane orientation with more $Co^{3+}$ exposed on surface that the nanostructured grains present, the more oxygen vacancies and $Co^{2+}$ ions in $O$ sites there will be, which is the mechanism of RTFM. The oxygen vacancies and lattice distortion caused by $Co^{2+}$ substitution in $O$ sites and other defects also accommodate $Li^+$, and hence increase the Li storage capacity. **(v)** Another possible reason for the worse Li-ion battery performance in smaller sized P270 - W300 samples comparing to that in P350 - W350 is due to the quantum electron refinement effect when the grain/particle size < 10nm, which decrease the electronic conductivity in smaller sized P270 - W300 samples.

We propose that the higher surface areas of the porous nanoplatelets, permits, on cycling, an increase in the extent of polymer layer formation per on the surface of active materials. The gradual increase of the excess capacity over the first 30 cycles is consistent with the proposal, that not all the surface layer is covered on the first discharge, or the internal surface within the pores are more difficult to access, so it appears that the polymer layer builds up slowly, over a number of cycles.

We note that the discharge capacity cycling performance of the P350 sample is nearly the same as that of the W350 at the 50 mA h/g rate, but the cycling performance of the W350 sample is significantly higher than those of the others when the rate increases up to 400 mA h/g, as shown in Fig. 4(a) and (b), which may imply enhancement of the mechanism behind the fast charge performance in the Li-ion battery.

The exposed surface of $Co_3O_4$ (110) may present two possible terminations in nanostructures[18, 71, 83, 95]. First principle calculation indicated that both terminations possible stable in nanostructures and possible lead ferromagnetism different from the one in the bulk[82, 95], the surface electronic states are present in the lower half of the bulk band gap and cause partial metallization of both surface terminations which consistent with the Scanning Tuning Microscope (STM) experiments on $Co_3O_4$ nanowires [96]. The exposed $Co_3O_4$ (110) surface electronic state direct dominate the electronic conductivity, chemical catalysis, electrochemical interface reactivity and further the Li-ion battery performance. In additional, the exposed $Co_3O_4$ (110) surface electronic states are responsible for the charge compensation / charge -mediated mechanism stabilizing polar terminations, which may also be the mechanism of room temperature ferromagnetism in the exposed $Co_3O_4$ (110) surface. The presented results in this paper all support above scenario.

## 4. Conclusions

In summary, in the $Co_3O_4$ cubic spinel nanostructures, improved {110} plane orientation is associated with more {110} planes exposed on the surface, a more dense atomic array, and a smaller lattice parameter, which directly lead to bulk conductivity improvement. More $Co^{3+}$ in an octahedral (*O*) position exposed on the surface is associated with more $Co^{3+}$ being changed to $Co^{2+}$ in the *O* position due to surface oxygen vacancies, which directly leads to surface room temperature ferromagnetism in the nanostructures and also leads to the significant improvement of surface catalysis, surface electronic conductivity, and finally, surface electrochemical reactivity, which is responsible for further improvement in the performance of lithium-ion batteries. We achieved a charge capacity that was nearly double the value of the theoretical capacity, a fast charge rate with long cycling life, and surface room temperature ferromagnetism for $Co_3O_4$ porous nanowires. These results show the importance of refinement control of the surface of nanostructures in the preparation of transition metal oxides as multifunctional nanomaterials.

**Acknowledgements**

The authors thank Dr. T. Silver for her help and useful discussions. This work is supported by the Australian Research Council through a Discovery project.

**References**

[1] A. Arico, P. G. Bruce, B. Scrosati, J. M. Tarascon, and W. V. Schalkwijk, *Nature Mater.,* **2005,** 4, 366.
[2] K. T. Nam, D.W. Kim, P. J. Yoo, et al., *Science,* **2006,** 312, 885-888.
[3] X. W. Lou, D. Deng, J. Y. Lee, et al., *Adv. Mat.,* **2008,** 20, 258.
[4] Y. G. Li, B. Tan, and Y. Y. Wu, *Nano Lett.,* **2008,** 8, 267.
[5] X. W. Xie, Y. Li, Z. Q. Liu, et al., *Nature,* **2009**, 458, 746.
[6] J. Feng and F. Heinz, *Angew. Chem. Int. Ed.,* **2009**, 48, 1841 .
[7] P. L. Meena, R. Kumar, C. L. Prajapat, K. Sreenivas, and V. Gupta, *J. Appl. Phys.* **2009,** 106, 024105.
[8] S. Takada, M. Fujii, S. Kohiki, T. Babasaki, H. Deguchi, M. Mitome, M. Oku, *Nano Lett.,* **2001,** 1, 379.
[9] K. Takada, H. Sakurai, E. Takayama-Muromachi, et al, *Nature,* **2003, 422** 53.
[10] B. Kang and G. Ceder, *Nature,* **2009,** 458, 190.
[11] B. E. Conway, *J. Electrochem. Soc.,* **1991,** 138, 1539.
[12] L. Tian, H. Zou, J. Fu, X. Yang, Y. Wang, et al, *Adv. Funct. Mater.* **2010**, 20, 617–623
[13] G. G. Amatucci, F. Badway, A. Du, and T. Zheng, *J. Electrochem. Soc.*, **2001,** 148, A930.
[14] J. M. Tarascon and M. Armand, *Nature,* **2001,** 414, 359.
[15] A. Gulino, G. Fiorito, and I. Fragal*, J. Mater. Chem.,* **2003,** 13, 861.
[16] S. Takada, M. Fujii, S. Kohiki, T. Babasaki, et al*, Nano Lett.,* **2001,** 1, 379.
[17] S. Kumagai, S. Yoshii, N. Matsukawa, K. Nishio, et al, *Appl. Phys. Lett.,* **2009**, *94, 083103.*
[18] J. Chen, A. Selloni, **2010**, *arXiv:*1201.6013.
[19] D. Mihailovic, et al., *Phys. Rev. Lett.*, **2003**, 90, 146401(2003).
[20] M. S. Seehra et al, *Phys. Rev. B,* **1979,** 19, 6620(1979).
[21] Burgardt et al., *Solid State Commun.* **1977**, 22, 153 (1977).]
[22] N.A. Chernova, G.M. Nolis, F.O. Omenya, H. Zhou, et al, **2011,** *J. Mater. Chem.* 21, 9865.
[23] G. D. Du, Z. P. Guo, S. Q. Wang, R. Zeng, et al, *Chemical Communications*, **2009,** 46, 1106.
[24] J. Liu, W.J. Wang, Z.P. Guo, R. Zeng, S.X. Dou, *Chem. Comm.***2010**, 46, 3887 (2010).
[25] R. Zeng, et al, unpublished works. ]

[26] H. G. Yang, C. H. Sun, S. Z. Qiao, J. Zou, G. Liu, et al., *Nature*, **2008,** 453, 638.
[27] R. Zeng, H. K. Liu, S. X. Dou, *Supercond. Sci. Technol.* **1998,** 11, 770.
[28] R. Zeng, J. Q. Wang, Z. X. Chen, W. X. Li, and S. X. Dou, *J. Appl. Phys.* **2011,** 109, 07B520.
[29] W. L. Yao, J. Yang, J. L. Wang, and Y. N. Nuli, *J. Electrochem. Soc.,* **2008,** 155, A903.
[30] W. Y. Li, L. N. Xu, and J. Chen, *Adv. Funct. Mater.,* **2005,** 15, 851..
[31] N. Du, H. Zhang, B. D. Chen, J. B. Wu, D. R. Yang, et al., *Adv. Mater.,* **2007,** 19, 4505.
[32] K. M. Shaju, F. Jiao, A. Debart, and P. G. Bruce, *Phys. Chem. Chem. Phys.,* **2007,** 9, 1837.
[33] X. W. Lou, D. Deng, J. Y. Lee, J. Feng, and L. A. Archer, *Adv. Mater.,* **2008**, 20, 258.
[34] Y. G. Li, B. Tan, and Y. Y. Wu, *Nano Lett.*, **2008**, 8, 265.
[35] B. Y. Geng, F. M. Zhan, C. H. Fang, and N. Yu, *J. Mater. Chem.,* **2008**, 18, 4977.
[36] X. Wang, X. Y. Chen, L. S. Gao, et al, *J. Phys. Chem. B*, **2004,** 108, 16401.
[37] D. Larcher, G. Sudant, J.B. Leriche, Y. Chabre, J.M. Tarascon, *J. Electrochem. Soc.* **2002,** 149 (2002) A234.]
[38] H.J. Liu, S.H. Bo, W.J. Cui, F. Li, C.X. Wang, Y.Y. Xia, *Electrochim. Acta,* **2008,** 53, 6497.
[39] X.L. Zhang, C.S.M. Lee, D.M.P. Mingos, D.O. Hayward, *Appl. Catal. A: Gen.* **2003,** 248, 129.
[40] V.Y. Bychkov, Y.P. Tyulenin, M.M. Slinko, V.N. Korchak, *Appl. Catal. A: Gen.* **2007,** 321, 180.
[41] J. Llorca, P.R.D.I. Piscina, J.A. Dalmon, N. Homs, *Chem. Mater.* **2004,** 16, 3573.
[42] L.B. Backman, A. Rautiainen, M. Lindblad, O. Jylh, A.O.I. Krause, *Appl. Catal. A* **2001,** 208 (2001) 223.]
[43] K.M. Shaju, F. Jiao, A. Debart, P.G. Bruce, *Phys. Chem. Chem. Phys.* **2007,** 9 (2007) 1837.
[44] J.M. Tarascon, S. Grugeon, M. Morcrette, S. Laruelle, P. Rozier, P. Poizot, *C. R. Chim.* **2005,** 8, 9;
[45] R. Dedryvere, S. Laruelle, S. Grugeon, P. Poizot, et al, *Chem. Mater.* **2004,** 16, 1056.]
[46] G. Gachot, S. Grugeon, M. Armand, S. Pilard, et al, *J. Power Sources* , 2008 178, 409.]
[47] J.Q. Wang, G.D. Du, R. Zeng, et al, *Electrochimica Acta,* **2010,** 55, 4805.
[48] J.Q. Wang, B. Niu, G.D. Du, R. Zeng, et al, *Mater. Chem. Phys.,* **2011,** 126, 747.
[49] P Zhang, Z.P. Guo, Y.D. Huang, et al, *J. Power Sources,* **2011,** 196, 6987.
[50] L. He, C. Chen, N.Wang,W. Zhou, and L. Guo, *J. Appl. Phys.* **2007,** 102, 103911.
[51] X. Batlle and A. Labarta, *J. Phys. D,* **2002,** 35, R15.
[52]. X. G. Zheng, C. N. Xu, K. Nishikubo, K. Nishiyama, et al, *Phys. Rev. B,* **2005,** 72, 014464.
[53] S.D. Tiwari and Rajeev, K.P., *Phys. Rev. B,* **2005,** 72, 104433.
[54] E. L. Salabas¸ et al, *Nano Lett.* **2006,** 6, 2977.
[55] E. Winkler, Zysler, Vasquez Mansilla, R. D. M., and Fiorani, D., *Phys. Rev. B,* **2005,** 72, 132409.
[56] J.B. Yi, J. Ding, Y.P. Feng, G.W. Peng, et al. *Phys. Rev. B,* **2007,** 76, 224402.
[57] S. Morup, and C. Frandsen, *Phys. Rev. Lett.* **2004,** 92, 217201.
[58] R. H. Kodama, S. A. Makhlouf, and A. E. Berkowitz, *Phys. Rev. Lett.* **1997,** 79, 1393.
[59] X. J. Yang, Y. Makita, Z. H. Liu, K. Sakane, and K. Ooi, *Chem. Mater.* **2004,** 16, 5581-5588.
[60] A. Tomou, D. Gournis, I. Panagiotopoulos, et al. *J. Appl. Phys.* **2006,** 99, 123915.
[61] A. Punnoose, and M. S. Seehra, *J. Appl. Phys*. 91, 7766 (2002).]
[62] H. G. Katzgraber, F. Pazmandi, C. R. Pike, K. Liu, et al, *Phys. Rev. Lett*. **2002,** 89, 257202.
[63] C. R. Pike, C. A. Ross, R. T. Scalettar, and G. Zimanyi, *Phys. Rev. B,* **2005,** 71, 134407.
[64] C. D. Spencer and D. Schroeer, *Phys. Rev. B,* **1974,** 9, 3658.
[65] T. Ambrose, and C. L. Chien, *Phys. Rev. Lett.,* **1996**, *76*, 1743.
[66] L. He, C. Chen, N. Wang, W. Zhou, and L. Guo, *J. Appl. Phys.,* **2007**, 102, 103911.
[67] E. L. Salabas, A. Rumplecker, F. Kleitz, F. Radu, and F. Schuth, *Nano Lett.,* **2006**, 6, 2977.
[68] M. J. Benitez, O. Petracic, E. L. Salabas, *Phys. Rev. Lett.,* **2008,** 101, 097206.
[69] J. P. Beaufils and Y. Barbaux, *J. Appl. Cryst.,* **1982,** 15, 301–307.
[70] J. Ziolkowski and Y. Barbaux, *J. Mol. Catal.,* **1991,** 67, 199–215.
[71] A. Serrano, F. Pinel, A. Quesada, et al., *Phys. Rev. B*, **2009,** 79, 144405.
[72] Y. Ikedo, J. Sugiyama, H. Nozaki, et al., *Phys. Rev. B*, **2007,** 75, 054424.
[73] N. Tristan, V. Zestrea, G. Behr, et al., *Phys. Rev. B*, **2008,** 77, 094412.
[74] M. A. Langell, J. G. Kim, D. L. Pugmire, and W. McCarroll, *J. Vac. Sci. Technol.* **2001**, A 19, 1977.
[75] D. Briggs and M. P. Seah, *Practical Surface Analysis* ~Wiley, New York, **1990**.
[76] B. A. Sexton, A. E. Hughes, and T. W. Turney, *J. Catal*. **1986,** 97, 390.
[77] J. G. Kim, D. L. Pugmire, D. Battaglia, and M. A. Langell, *Appl. Surf. Sci*. **2000,** 165, 70.
[78] Z. X. Shen *et al.*, *Phys. Rev. B,* **1990,** 42, 1817.
[79] G. A. Carson, M. H. Nassir, and M. A. Langell. *J. Vac. Sci. Technol*. **1996,** A 14, 1637.

[80] J. van Elp, J. L. Wieland, H. Eskes, P. Kuiper, G. A. Sawatasky, et al, *Phys. Rev. B,* **1991,** 44, 6090.
[81] N. S. McIntyre and M. G. Cook, *Anal. Chem*. **1975,** 47, 2208.
[82] M. A. Langell, G. A. Carson, M. Anderson, L. Peng, and S. Smith, *Phys. Rev. B* **1999,** 59, 4791.
[83] S. C. Petitto and M. A. Langell, *J. Vac. Sci. Technol*. **2004,** A 22.4. 1690.
[84] R. Xu and H. C. Zeng, *Langmuir* **2004,** *20,* 9780-9790
[85] T. J. Chuang, C. R. Brundle, and D. W. Rice, *Surf. Sci.* **1976,** 59, 413..
[86] V. M. Jimenez, A. Fernadez, J. P. Espinos, et al*, J. Electron Spectrosc. Relat. Phenom*. **1995,** 71, 61.
[87] M. Oku and Y. Sato, *Appl. Surf. Sci*. **1992,** 55, 37.
[88] N. E. Rajeevan, P. P. Pradyumnan, R. Kumar, et al, *Appl. Phys. Lett.,* **2008,** 92, 102910.
[89] Y. Ikedo, J. Sugiyama, H. Nozaki, and H. Itahara, *Phys. Rev. B***, 2007,** 75, 054424.
[90] W. L. Roth, *J. Phys. Chem. Solids***, 1964,** 25, 1.
[91] N. Kita, N. Shibuichi, and S. Sasaki, *J. Synchrotron Rad.*, **2001,** 8, 446-448.
[92] Y. N. Sharma, G. V. Subba Rao, and B. V. R. *Solid State Ionics,* **2008,** 179, 587–597.
[93] P. Poizot, S. Laruelle, S. Grugeon, L. Dupont, and J.-M. Tarascon, *Nature,* **2000,** 407, 28.
[94] F. Zhan, B. Geng, and Y. Guo. *Chem. Eur. J.,* **2009,** 15, 6169 – 6174.
[95] J. Chen, X. Wu, and A. Selloni, *Phys. Rev. B*, **2011,** 83, 245204.
[96] Y Sun, J Yang, R Xu, L He, R Dou, and J Nie, *Applied Physics Letters,* **2010,** 96, 262106.

# Figure Captions

**Figure 1.** (Color online). TEM images of selected $Co_3O_4$ nanostructures: (a) P350, (b) P500, (c) W300, and (d) W350. The insets are the SAED patterns of the corresponding samples.

**Figure 2.** (Color online). HRTEM images of selected $Co_3O_4$ nanostructures: (a) P350, (b) P500, (c) W300, and (d) W350. The insets in (b) are the SAED patterns of the corresponding particle edge and centre. The insets in (d) indicate the sources of the enlarged HRTEM images.

**Figure 3.** (Color online). The {110} plane orientation (percentages of texture) and lattice parameter versus calcining temperature of $Co_3O_4$ nanostructured samples.

**Figure 4.** (Color online). Discharge rate capability and capacity retention for $Co_3O_4$ nanostructures at 50 mAh/g (a) and at 400 mAh/g (b).

**Figure 5.** Discharge/charge voltage profiles for W350 and W500 measured in the voltage range of 0.01 to 3.0V at a rate of 50mA/g.

**Figure 6.** (Color online). M vs. T curves under zero-field cooling (ZFC) and after field cooling (FC) at a field of 1 kOe for (a) nanoplatlets and (b) nanowires; the M vs. H loops of $Co_3O_4$ nanostructures at 5 K magnetisation versus applied field (M – H) loops are shown in (c) for both nanoplatlets and nanowires samples, the inset showing M – H loops at 305 K for the W350 and P270 samples.

**Figure 7.** (Color online). Summary of various properties vs. annealing temperature for the samples: the average grain size (a), the Néel temperature ($T_N$) (b) the coercive field ($H_C$) (c), and the magnetisation hysteresis loop horizontal shift ($H_{HS}$) under 1 kOe and 50 kOe FC (d).

**Figure 8**. (Color online). (A) Plots of $\chi T$ -T curves for selected $Co_3O_4$ nanostructures samples under field cooling (FCin an external magnetic field $\mu_0 H$ = 1 kOe; (B) Plots the results of $\Theta_{CW}$ (a), $\mu_{eff}$ (b), $\chi_0$ (c) and $\chi_1$ (d) after fitting the $\chi$-T curves under the $\mu_0 H$ = 1 kOe applied fields according to the equation (1) as descripted in text ; (C) Isothermal M-H hysteresis loops ($\mu_0 H$ < 1 T) for P270 nanopatelets sample at 305 K, it shows the ferromagnetic loops (blue line) with saturate moment $M_s$ = 0.014 emu/g and coercive field $\mu_0 H_C$ = 76 Oe after subtraction of the linear Pauli-like paramagnetic component (red line); (D) Isothermal M-H hysteresis loops ($\mu_0 H$ < 1 T) for W350 nanowires sample at 305 K, it shows the ferromagnetic loops (blue line) with saturate moment $M_s$ = 1.01 emu/g and coercive field $\mu_0 H_C$ = 126 Oe after subtraction of the linear Pauli-like paramagnetic component (red line).

**Figure 9.** (Color online). Conductivity Arrhenius plots for W350 and P350 samples.

**Figure 10.** (Color online). (a) Selected XPS spectra of $Co_3O_4$ nanostructures, (b) enlargement of Co-2P1/2 and Co-2P3/2 peaks, and (c) enlargement of O-1S peak.

**Figure 11.** (Color online). Crystal structure of $Co_3O_4$ sample with only Co ions shown (a); part of cell lattice structure with $O^{2-}$ shown (b), and after one oxygen ion is lost (c); the surface atomic configurations in the {111} (d), {110} (e), and {110} (f) planes after oxygen vacancy has caused surface octahedral $Co^{3+}$ to alter to $Co^{3+}|^{\bullet}$.

**Figure 12.** (Color online). TEM images of W350 sample after charge-discharge for (a) 20 cycles at 50 mAh/g rate, (b) 60 cycles at 400 mAh/g rate, (c) 100 cycles at 400 mA h/g rate. (d), (e) and (f) are the HRTEM images of the above (a), (b), and (c) samples, respectively. The circles and fringe line marks indicate active carbon and graphene cover layers, and the insets in (a), (b), and (c) show the corresponding SAED patterns.

# Figures

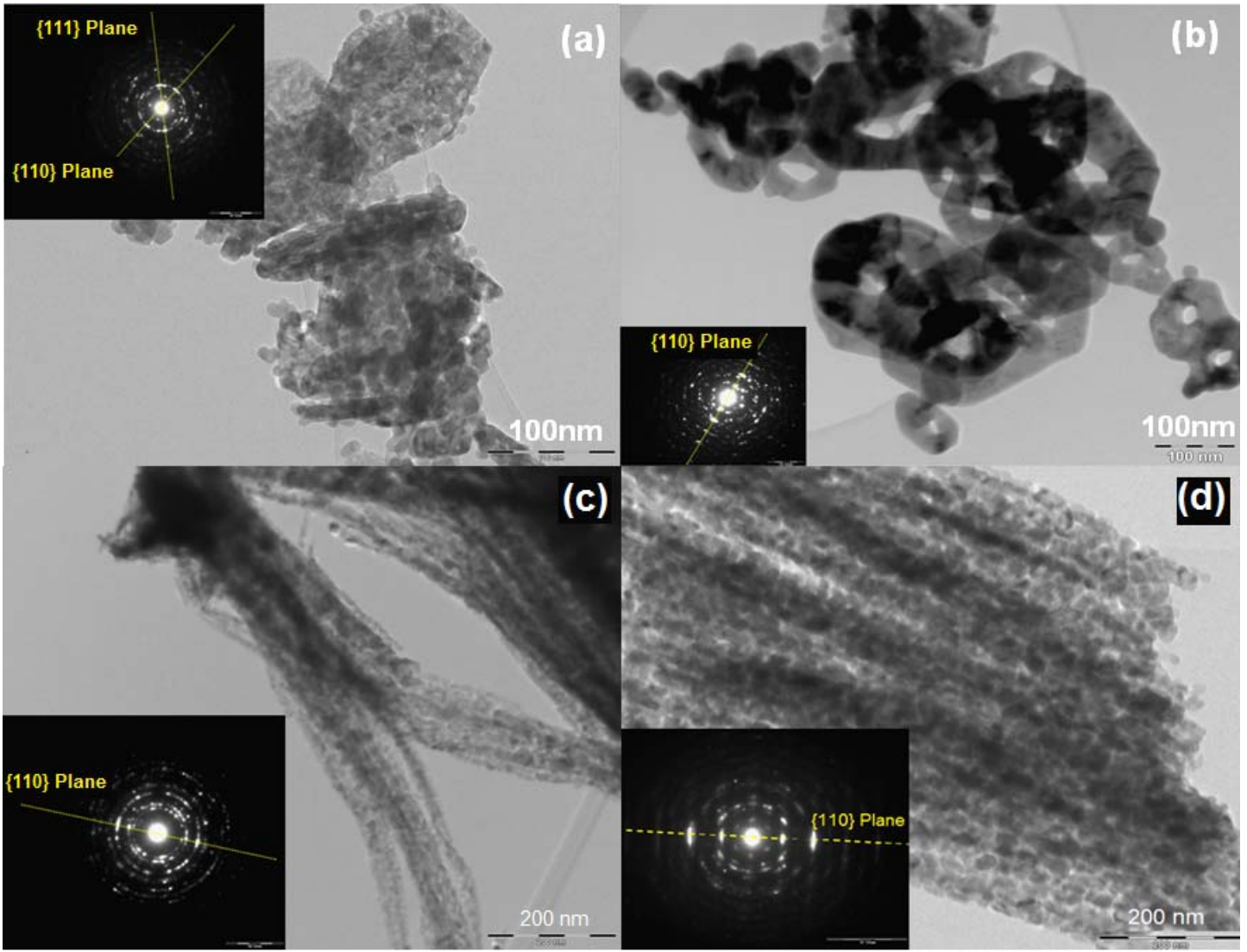

**Figure 1.** R. Zeng, et al.

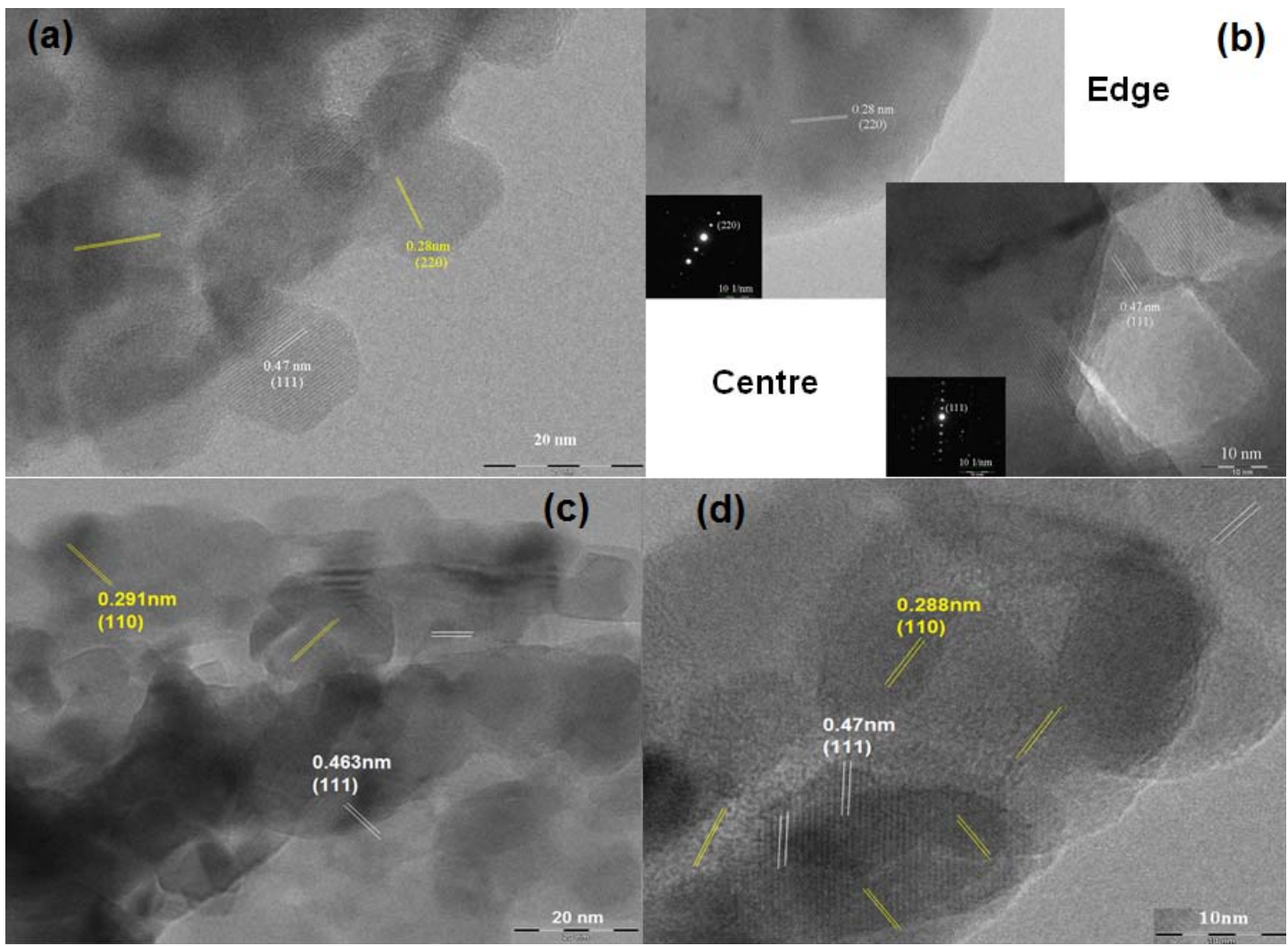

**Figure 2. R. Zeng, et al.**

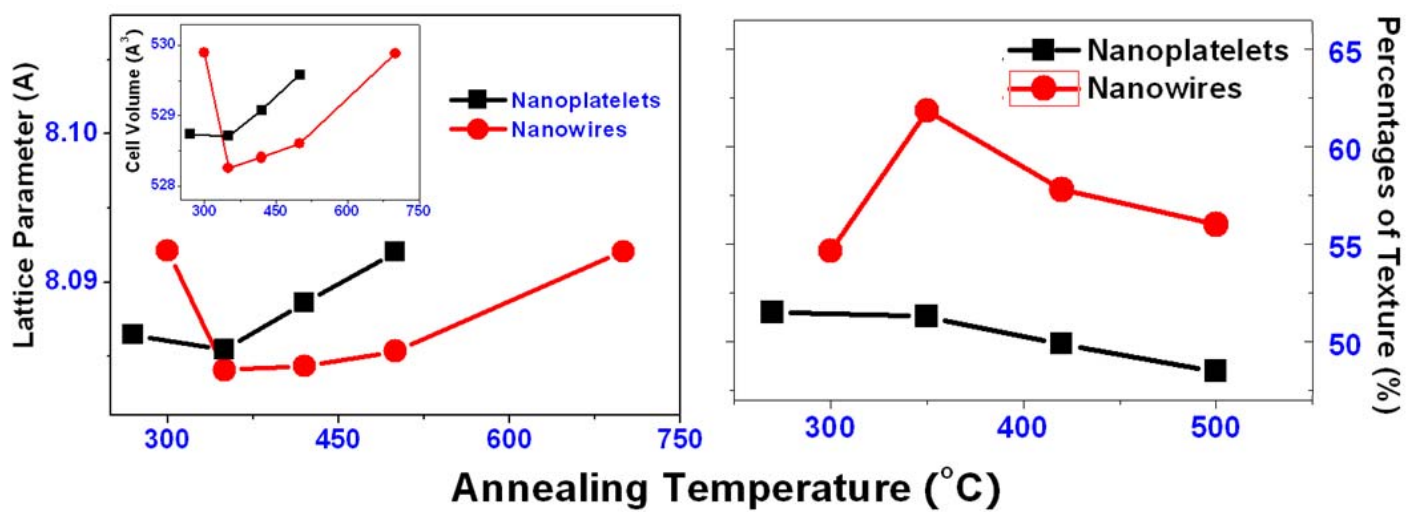

**Figure 3. R. Zeng, et al.**

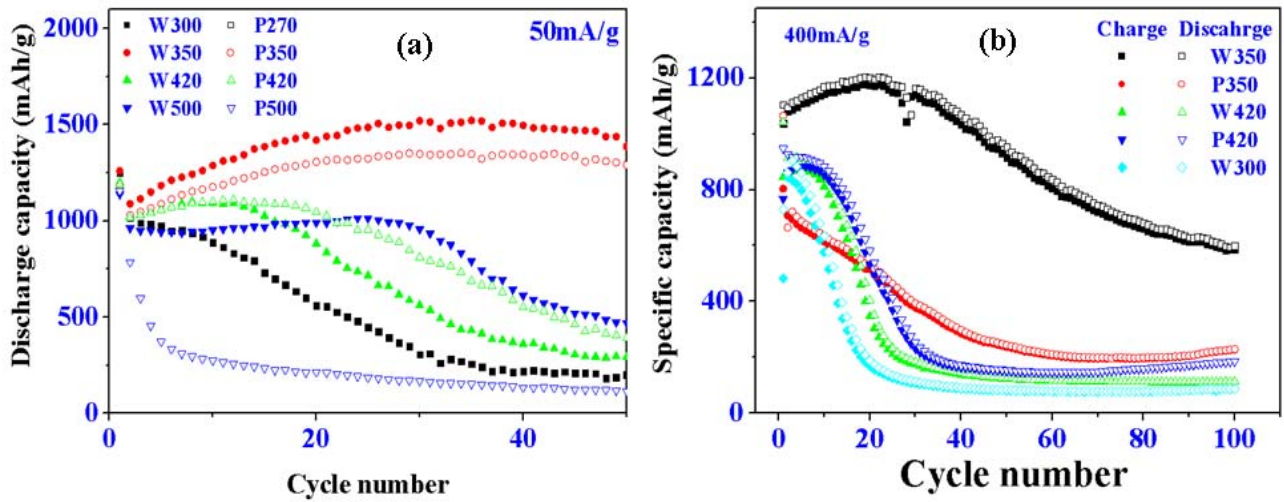

**Figure 4.** R. Zeng, et al.

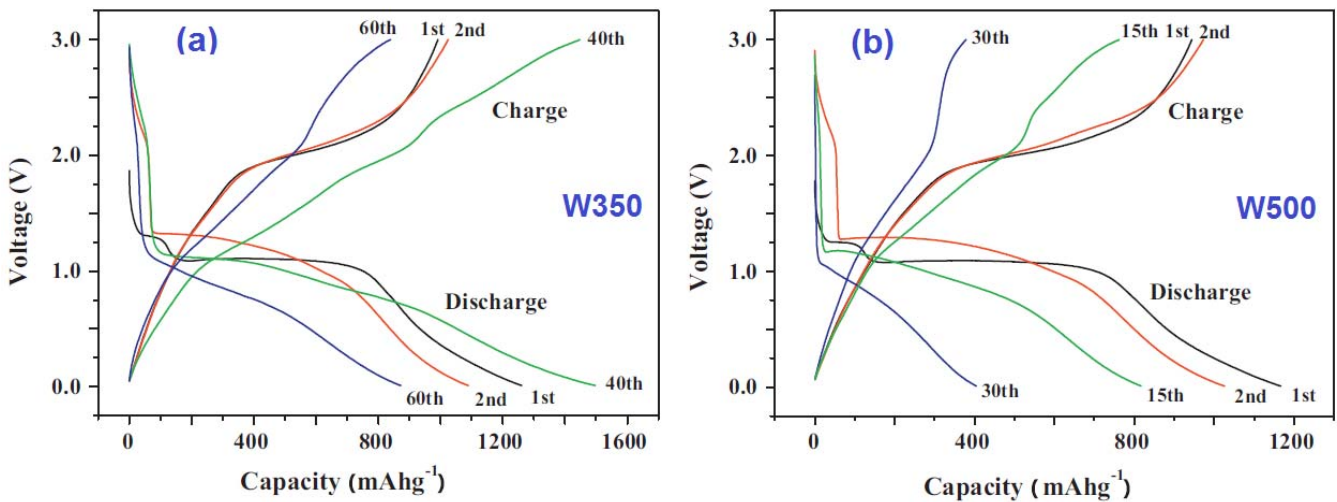

**Figure 5.** R. Zeng, et al.

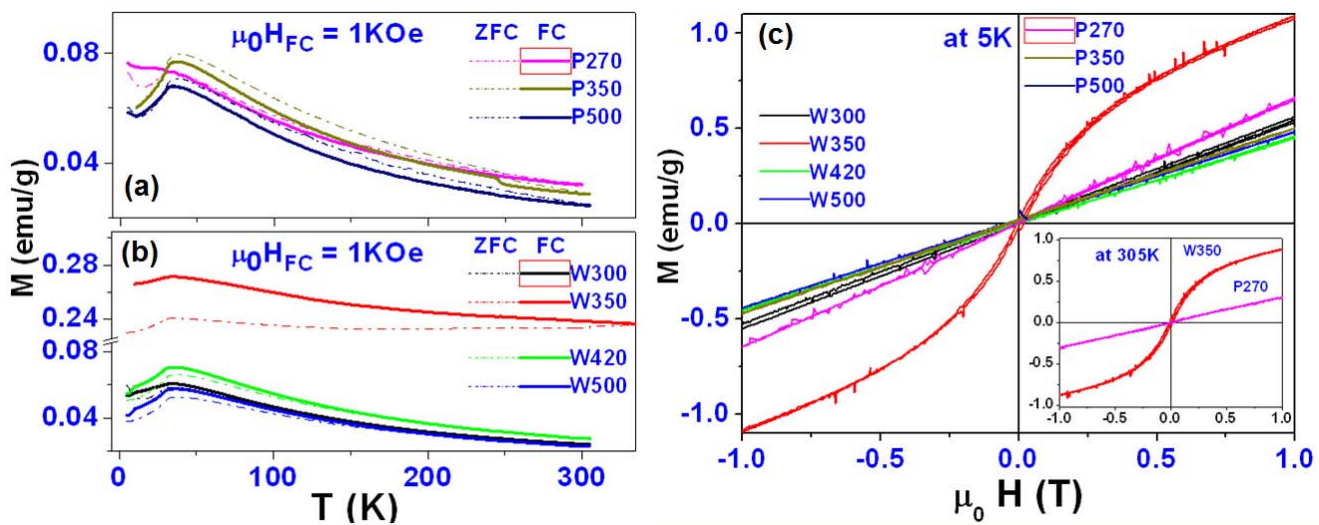

**Figure 6.** R. Zeng, et al.

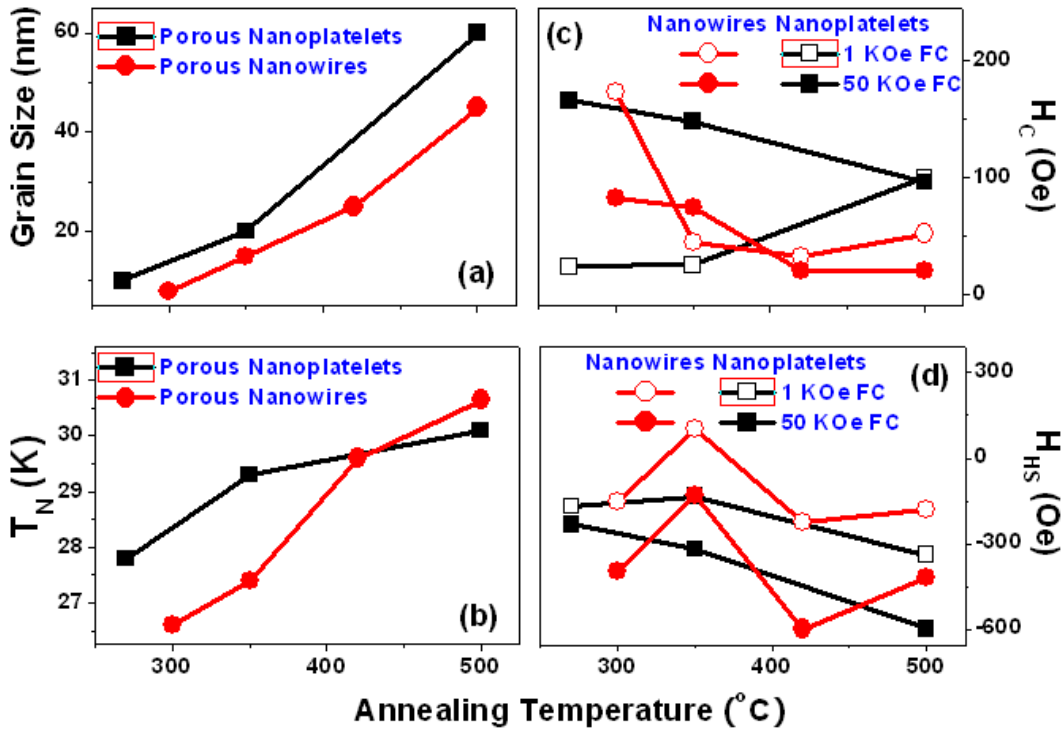

**Figure 7. R. Zeng, et al.**

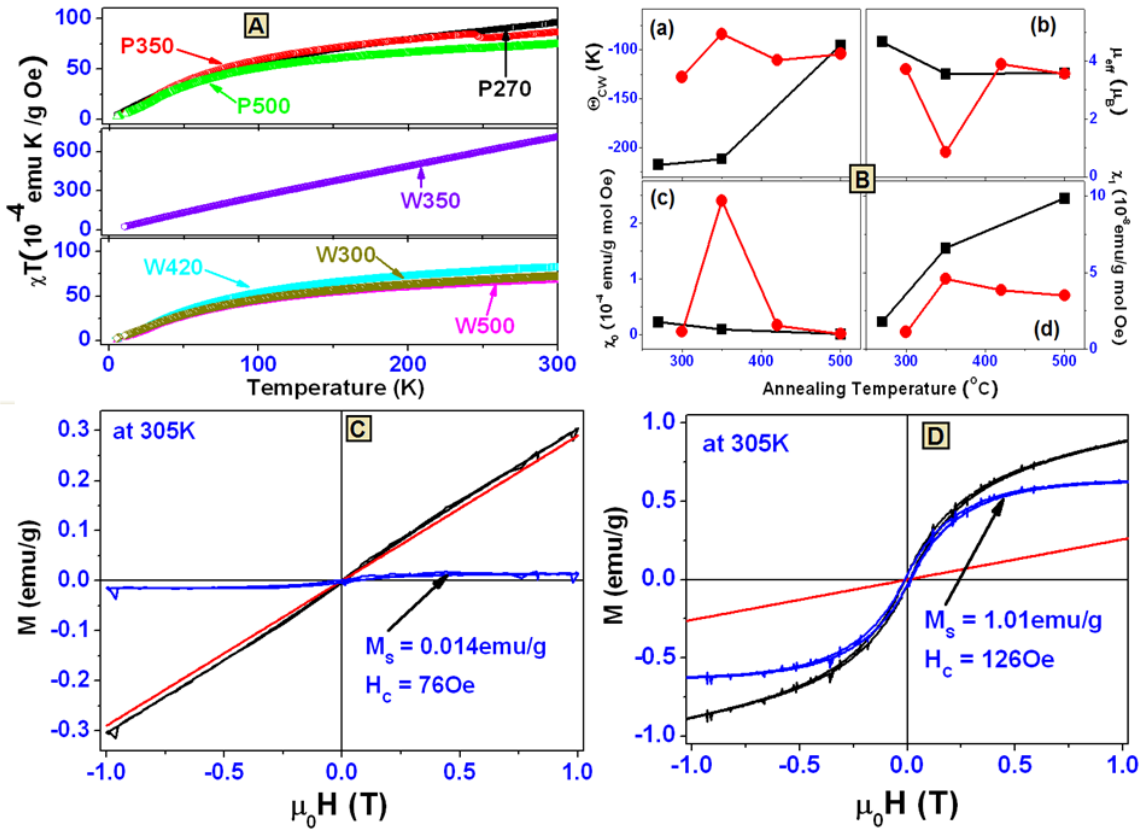

**Figure 8. R. Zeng, et al.**

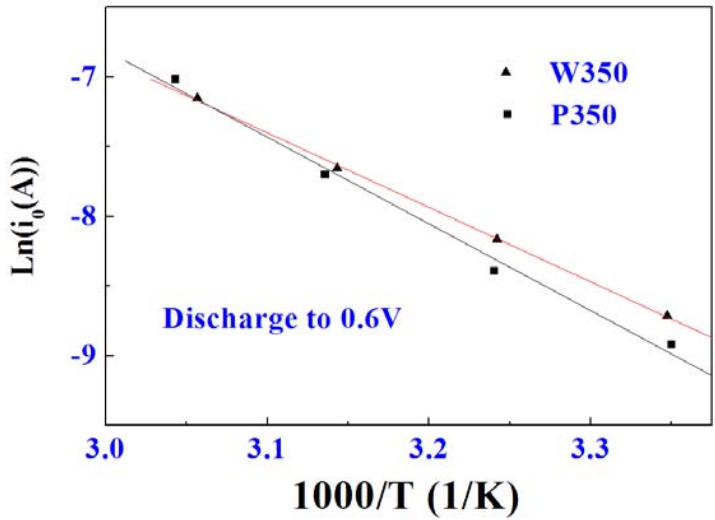

**Figure 9. R. Zeng, et al.**

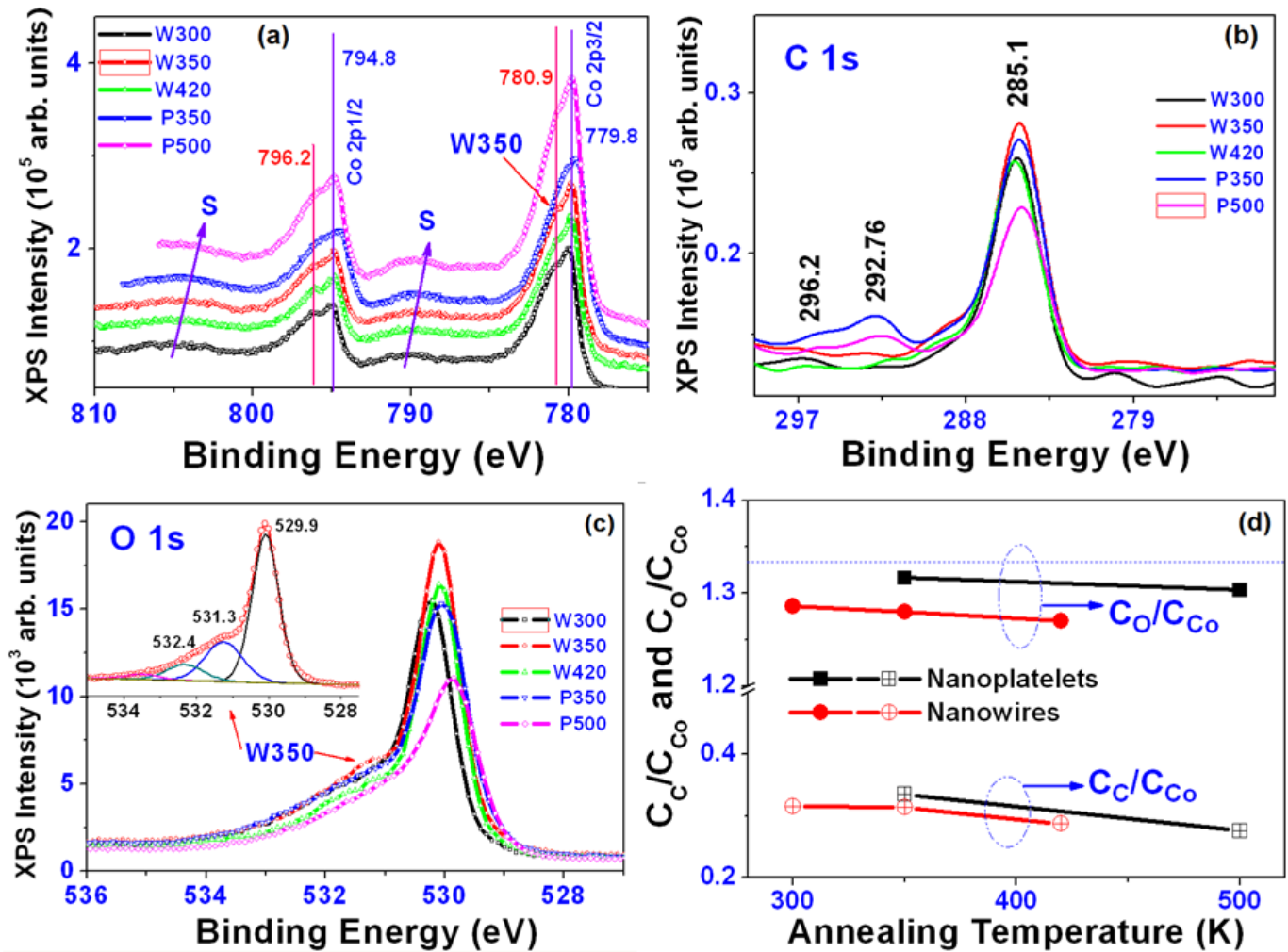

**Figure 10. R. Zeng, et al.**

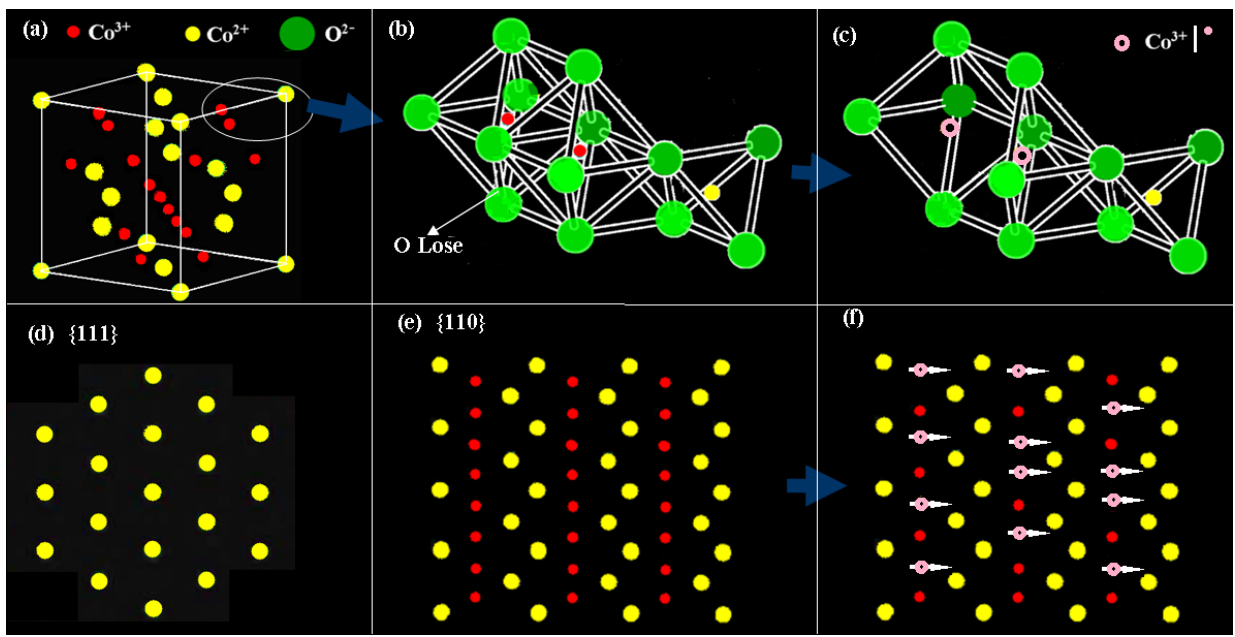

Figure 11. R. Zeng, et al.

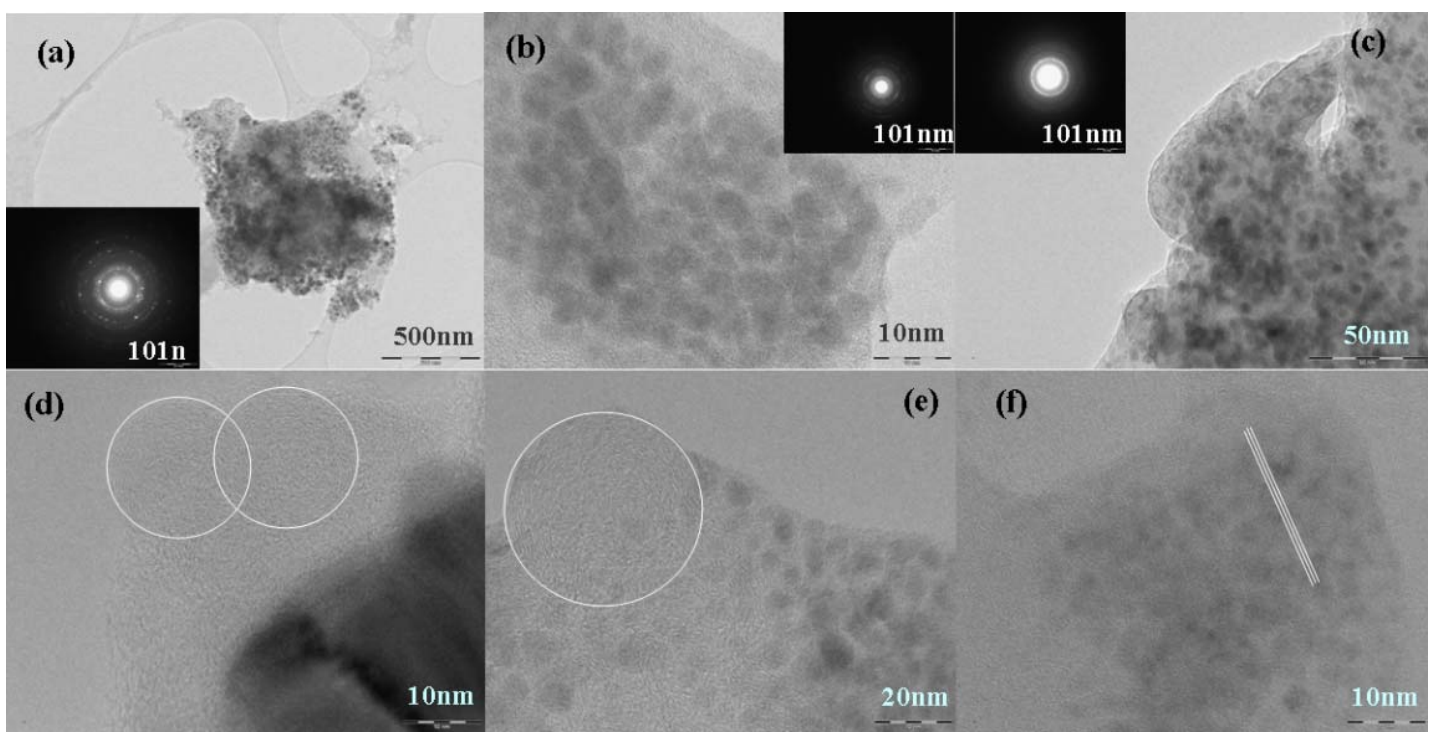

Figure 12. R. Zeng, et al.

# Supporting Information
# for

# {110} Plane Orientation Driven Superior Li-ion Battery Performance and Room Temperature Ferromagnetism in $Co_3O_4$ Nanostructures

R. Zeng[†,‡], J.Q. Wang[†,∥], G.D. Du[†,‡], W.X. Li[†], Z.X. Chen[†], S. Li[§], Z.P. Guo[†], S.X. Dou[†]

[†]*Institute for Superconducting and Electronic Materials, School of Mechanical, Materials & Mechatronics Engineering, University of Wollongong, NSW 2522, Australia.*

[‡]*Solar Energy Technologies, School of Computing, Engineering and Mathematics, University of Western Sydney, Penrith Sout, Sydney, NSW 2751, Australia*

[∥]*School of Materials Science and Engineering, University of Jinan, Jinan 250022, P. R. China.*

[§]*School of Materials Science and Engineering, University of New South Wales, Sydney NSW 2502, Australia.*

Address for Correspondence:

R. Zeng

Solar Energy Technologies
School of Computing, Engineering and Mathematics
University of Western Sydney
Penrith Sout, Sydney, NSW 2751, Australia
Electronic mail: r.zeng@uws.edu.au

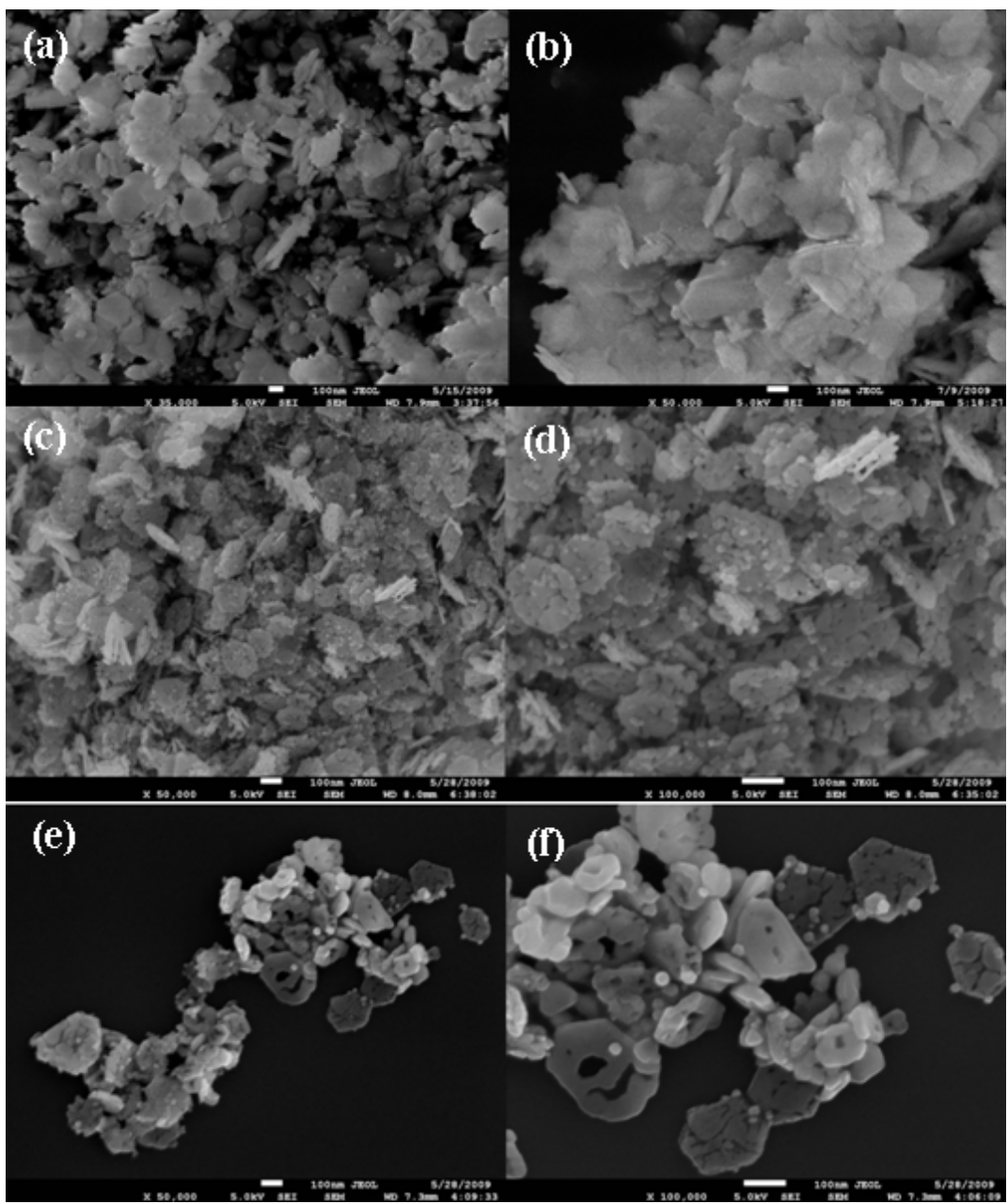

**Fig. S1.** FESEM images of selected β-Co(OH)$_2$ and Co$_3$O$_4$ nanoplatelets: (a) (b) β-Co(OH)$_2$; (c) (d) porous Co$_3$O$_4$ nanoplatelets prepared at 350°C; (e) (f) Co$_3$O$_4$ nanorings prepared at 500°C.

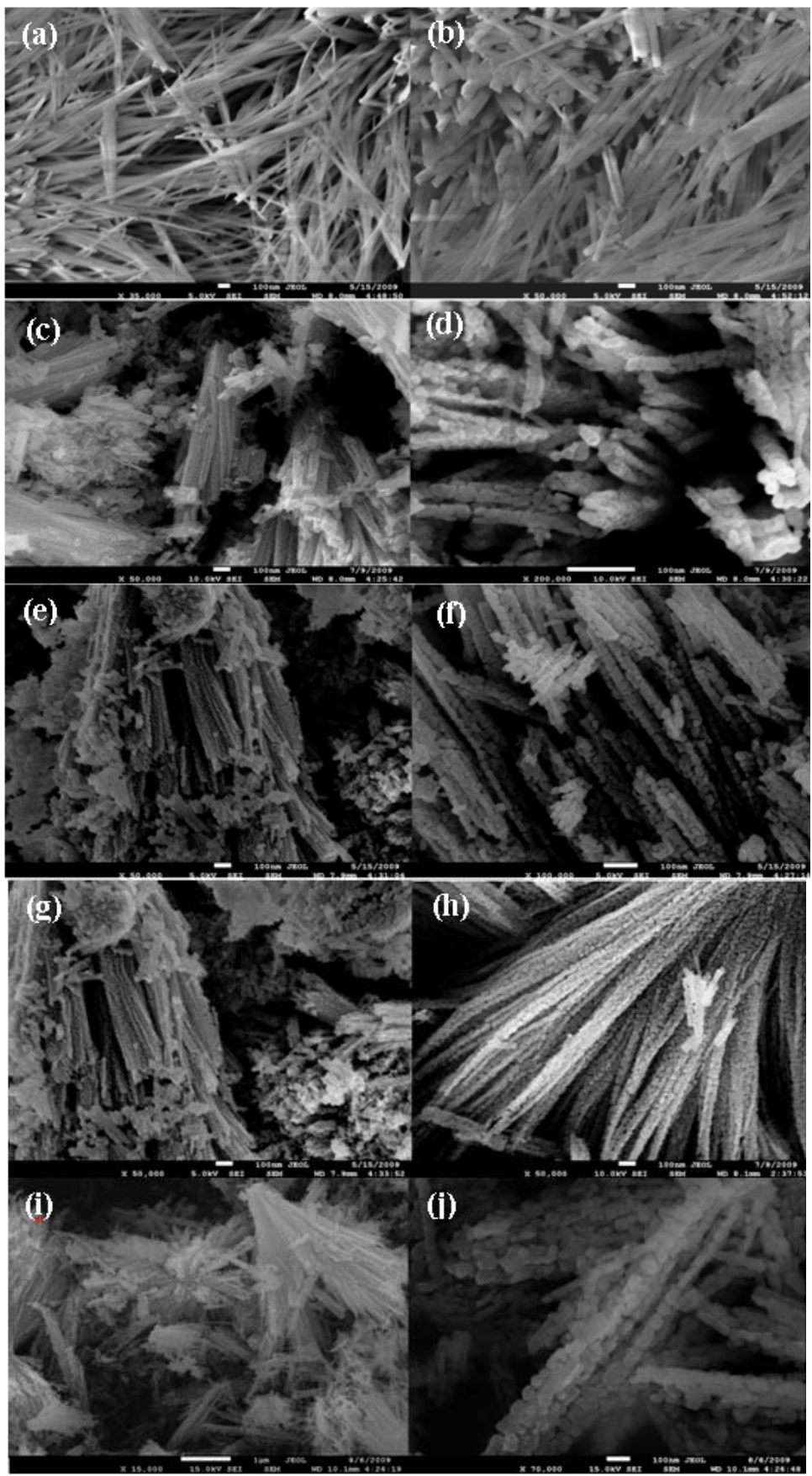

**Fig. S2.** FESEM images of selected β-Co(OH)$_2$ and Co$_3$O$_4$ nanowires: (a) (b) β-Co(OH)$_2$, (c) (d) W300, (e) (f) W350, (g) (h) W420, and (i) (j) W500.

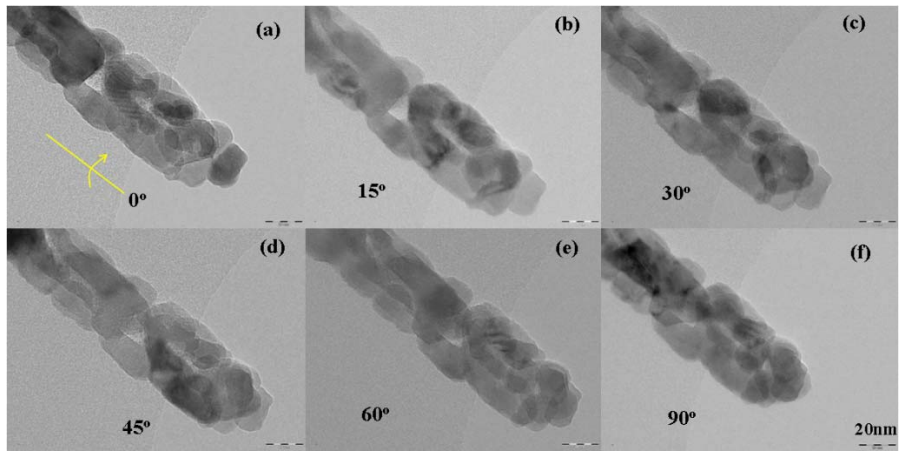

**Fig. S3.** TEM images of individual porous nanowire of W350 sample under axis turning of 0° (a), 15° (b), 30° (c), 45° (d), 60° (e), and 90° (f).

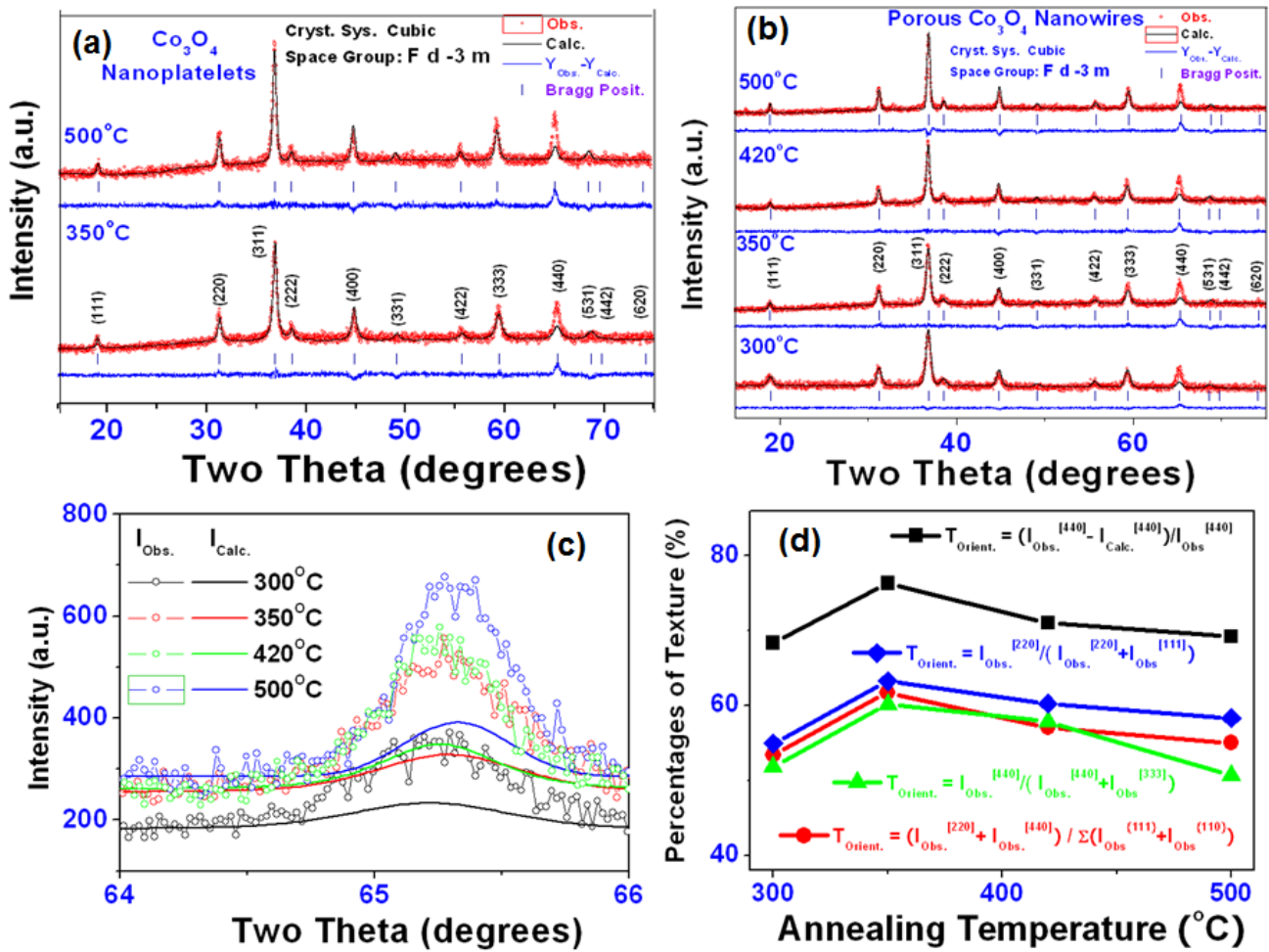

**Fig. S4.** XRD results for porous nanostructure samples sintered at different temperatures: **(a)** XRD patterns and Fullprof refinement calculation results for nanoplatelets; **(b)** XRD patterns and Fullprof refinement calculation results for nanowires; **(c)** XRD patterns of [440] peaks and Fullprof refinement calculation results; **(c)** the calculation results for {110} plane textures by four different methods, with the four formulas indicated on the figure.

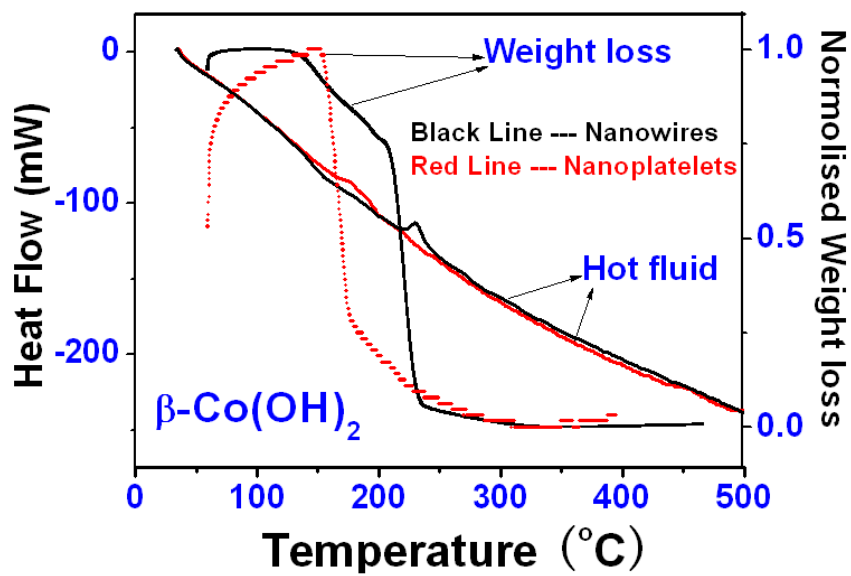

**Fig. S5.** TG and DSC curves of the β-Co(OH)$_2$ nanoplatelet and nanowire precursors.

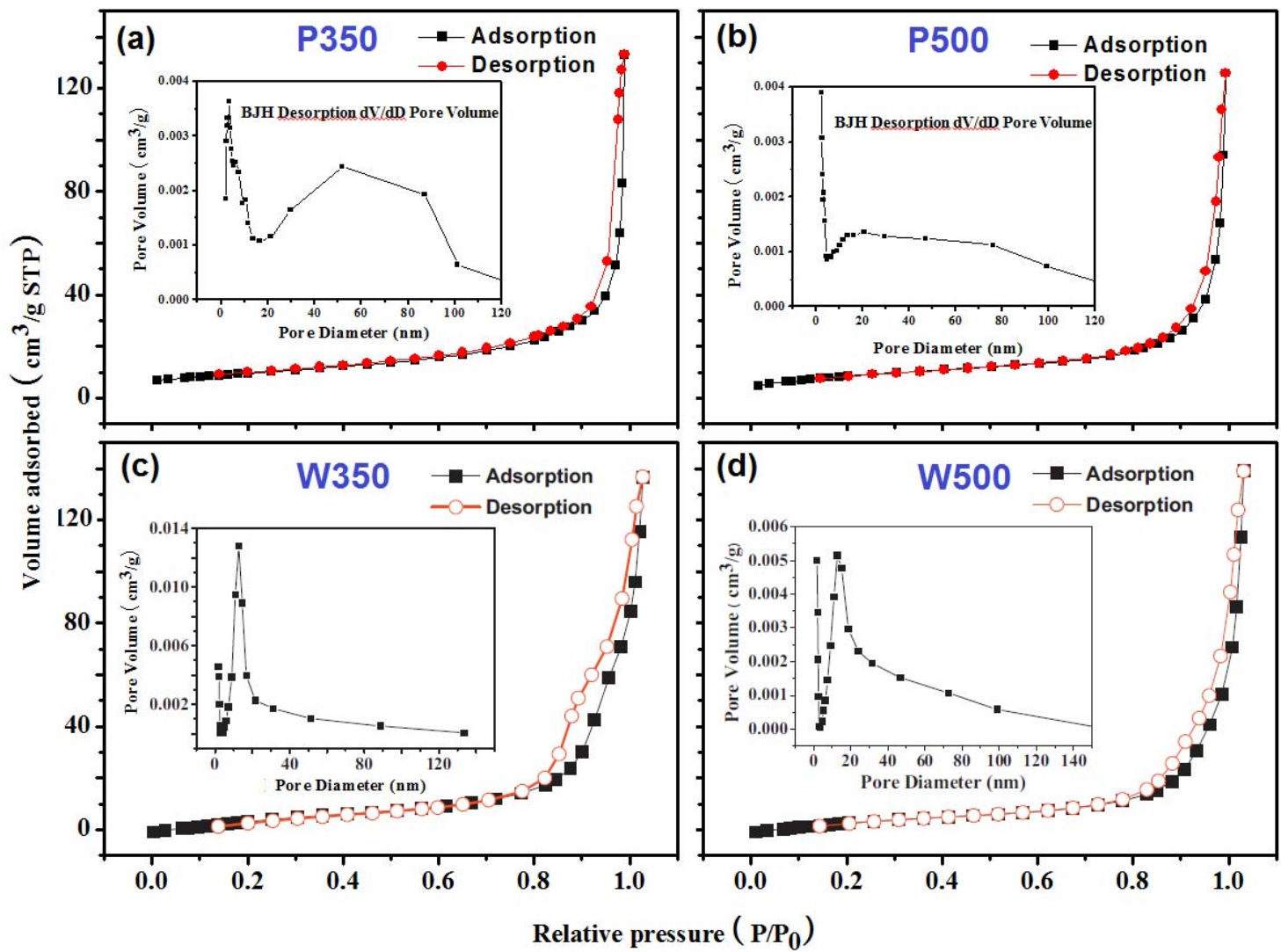

**Fig. S6.** N$_2$ adsorption/desorption isotherms and corresponding BJH pore size distributions (insets) for sample (a) P350, (b) P500, (c) W350 and (d) W500.

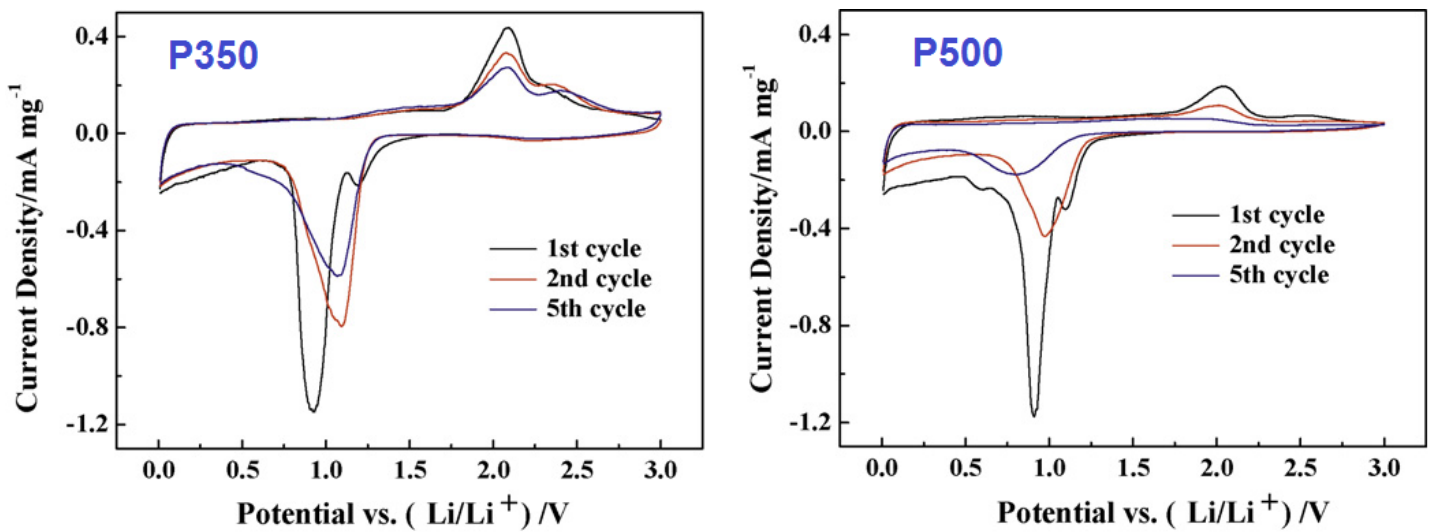

**Fig. S7.** Cyclic voltammogram profiles of electrodes made by P350 and P500 for 1st, 2nd and 5th cycles at a scan rate of $0.1\,\mathrm{mVs^{-1}}$.

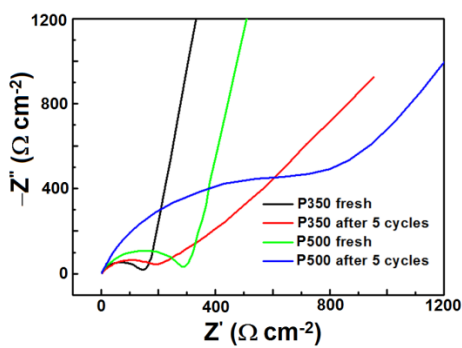

**Fig. S8.** Electrochemical impedance spectrum of the fresh $Co_3O_4$ nanoplatelets P350 and P500 electrodes and corresponding ones after 5 cycles.

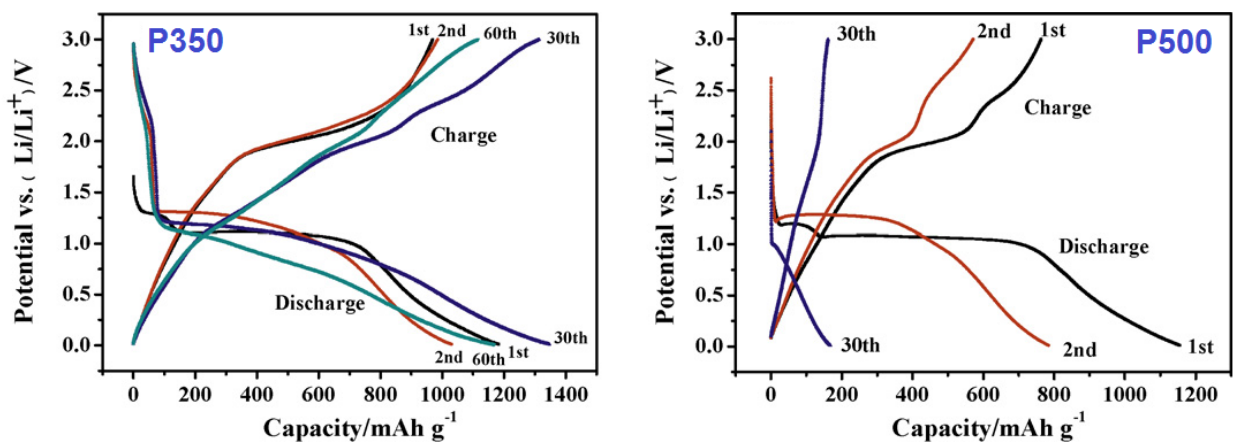

**Fig. S9.** Discharge/charge voltage profiles for P350 and P500 measured in the voltage range 0.01–3.0V at a rate of $50\,\mathrm{mAg^{-1}}$.

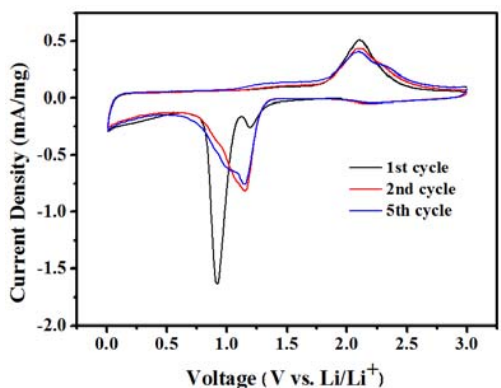

**Fig. S10.** Cyclic voltammogram profiles of electrodes made by W350 sample for 1st, 2nd and 5th cycles at a scan rate of 0.1 mV s$^{-1}$.

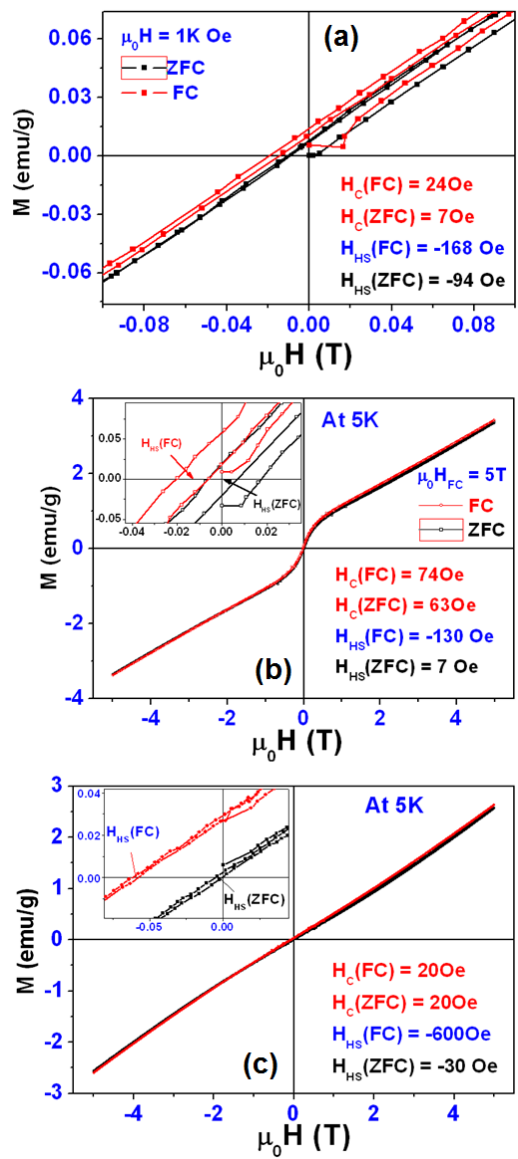

**Fig. S11.** The hysteresis loops measured with applied 5 T fields at 5 K after ZFC or FC for P270 (a), W350 (b) and W420 (c). Two insets in (b) and (c) show enlargement of low field range, the the horizontal shift $H_{HS}$ behaviours after ZFC and FC are indicated in the inserts, and the values of coercive fields $H_C$ and $H_{HS}$ are shown in the figures.

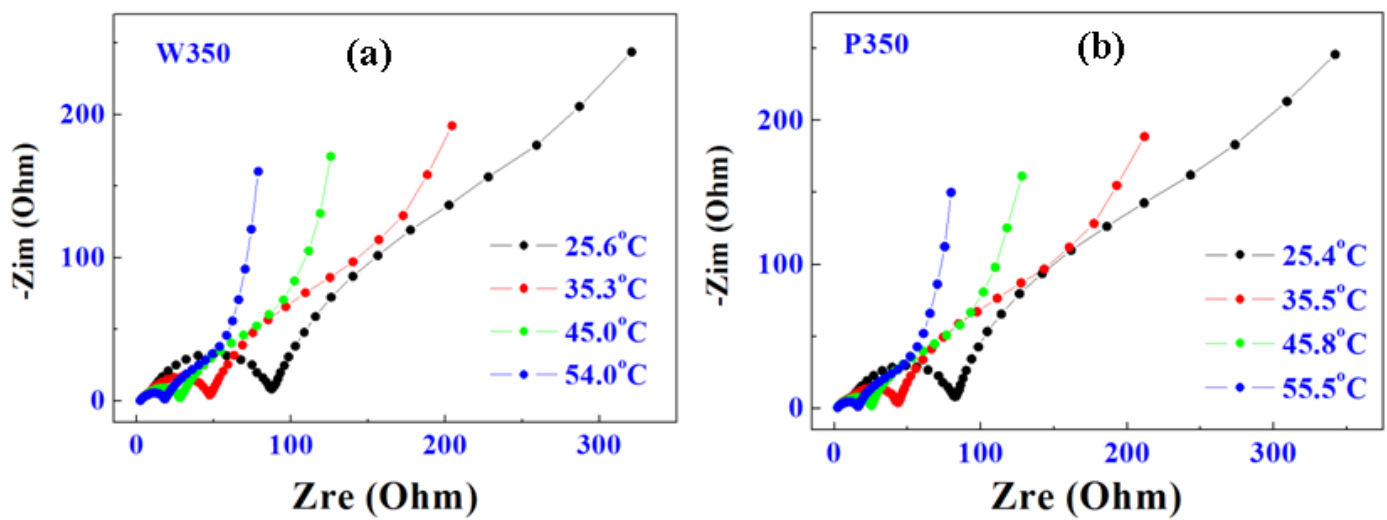

**Fig. S12.** Electrochemical impedance spectra with discharge to 0.6 V: (a) W350 and (b) P350 $Co_3O_4$ electrodes at different temperatures.

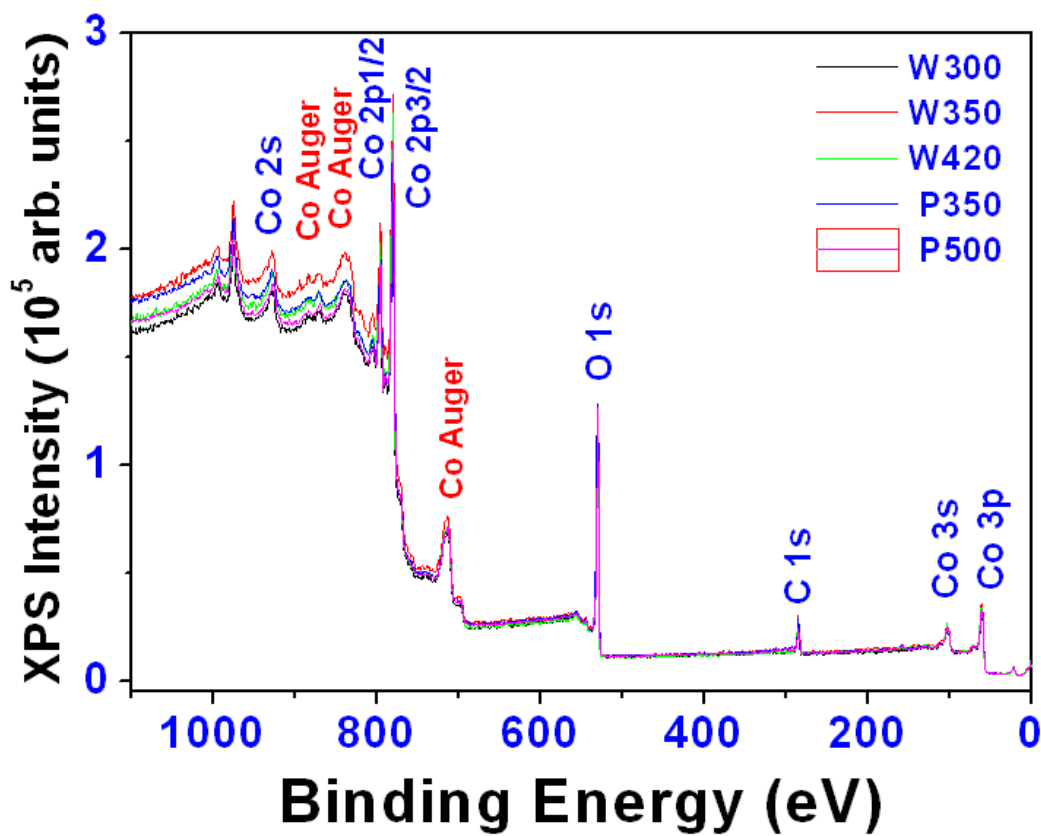

**Fig. S13.** Whole range of peaks XPS spectra of selected $Co_3O_4$ nanostructures.